\PassOptionsToPackage{table,xcdraw}{xcolor}
\documentclass[runningheads]{llncs}

\usepackage[misc]{ifsym}
\usepackage{array}
\usepackage{acro}
\usepackage{amsmath}
\usepackage{amsfonts}
\usepackage{graphicx}
\usepackage{caption}
\usepackage{makecell}
\usepackage{float}
\usepackage{footnote}
\usepackage{tabularx}
\usepackage{tikz}
\usepackage{placeins}
\usepackage[inline]{enumitem}
\usepackage{booktabs}
\usepackage{csquotes}
\usepackage{wasysym}
\usepackage{multirow}
\usepackage{pdflscape}
\usepackage{multicol}
\usepackage{arydshln}
\usepackage{adjustbox}
\usepackage{subcaption}
\usepackage{xspace}
\usepackage{algorithm}
\usepackage{algpseudocode}
\usepackage{hyperref}
\usepackage{longtable}
\usepackage{multicol}
\usepackage{hhline}
\usepackage{setspace}
\usepackage{bm}
\usepackage[T1]{fontenc}
\usepackage{listings}
\usepackage{xurl}
\usepackage{orcidlink}

\usepackage{tikz}
\newcommand\encircle[1]{%
  \tikz[baseline=(X.base)] 
    \node (X) [draw, shape=circle, inner sep=0.1, fill=black, text=white, font=\scriptsize] {\strut #1};%
}

\hypersetup{
colorlinks=true,
linkcolor=blue,
filecolor=magenta,      
urlcolor=cyan,
pdftitle={Overleaf Example},
pdfpagemode=FullScreen,
}

\setlist[enumerate]{
leftmargin=*,
}

\makeatletter
\renewcommand{\sectionautorefname}{\S\@gobble}
\renewcommand{\subsectionautorefname}{\S\@gobble}
\renewcommand{\subsubsectionautorefname}{\S\@gobble}

\makeatother

\newlist{researchquestions}{enumerate}{1}
\setlist[researchquestions]{label*=\textnormal{RQ\arabic*}}

\definecolor{verylightgray}{rgb}{.97,.97,.97}

\lstdefinelanguage{Solidity}{
	keywords=[1]{anonymous, assembly, assert, balance, break, call, callcode, case, catch, class, constant, continue, constructor, contract, debugger, default, delegatecall, delete, do, else, emit, event, experimental, export, external, false, finally, for, function, gas, if, implements, import, in, indexed, instanceof, interface, internal, is, length, library, log0, log1, log2, log3, log4, memory, modifier, new, payable, pragma, private, protected, public, pure, push, require, return, returns, revert, selfdestruct, send, solidity, storage, struct, suicide, super, switch, then, this, throw, transfer, true, try, typeof, using, value, view, while, with, addmod, ecrecover, keccak256, mulmod, ripemd160, sha256, sha3}, 
	keywordstyle=[1]\color{blue}\bfseries,
	keywords=[2]{address, bool, byte, bytes, bytes1, bytes2, bytes3, bytes4, bytes5, bytes6, bytes7, bytes8, bytes9, bytes10, bytes11, bytes12, bytes13, bytes14, bytes15, bytes16, bytes17, bytes18, bytes19, bytes20, bytes21, bytes22, bytes23, bytes24, bytes25, bytes26, bytes27, bytes28, bytes29, bytes30, bytes31, bytes32, enum, int, int8, int16, int24, int32, int40, int48, int56, int64, int72, int80, int88, int96, int104, int112, int120, int128, int136, int144, int152, int160, int168, int176, int184, int192, int200, int208, int216, int224, int232, int240, int248, int256, mapping, string, uint, uint8, uint16, uint24, uint32, uint40, uint48, uint56, uint64, uint72, uint80, uint88, uint96, uint104, uint112, uint120, uint128, uint136, uint144, uint152, uint160, uint168, uint176, uint184, uint192, uint200, uint208, uint216, uint224, uint232, uint240, uint248, uint256, var, void, ether, finney, szabo, wei, days, hours, minutes, seconds, weeks, years},	
	keywordstyle=[2]\color{teal}\bfseries,
	keywords=[3]{block, blockhash, coinbase, difficulty, gaslimit, number, timestamp, msg, data, gas, sender, sig, value, now, tx, gasprice, origin},	
	keywordstyle=[3]\color{violet}\bfseries,
	identifierstyle=\color{black},
	sensitive=true,
	comment=[l]{//},
	morecomment=[s]{/*}{*/},
	commentstyle=\color{gray}\ttfamily,
	stringstyle=\color{red}\ttfamily,
	morestring=[b]',
	morestring=[b]"
}

\newcommand*\halfcirc[1][1ex]{%
  \raisebox{-0.6ex}{%
    \begin{tikzpicture}
      \draw[fill] (0,0)-- (90:#1) arc (90:270:#1) -- cycle ;
      \draw (0,0) circle (#1);
    \end{tikzpicture}%
  }%
}
\DeclareAcronym{defi}{
    short = DeFi,
    long = decentralized finance
}

\DeclareAcronym{tvl}{
    short = TVL,
    long = total value locked
}

\DeclareAcronym{tvr}{
    short = TVR,
    long = total value redeemable
}

\DeclareAcronym{lp}{
    short = LP,
    long = liquidity provider
}

\DeclareAcronym{cdp}{
    short = CDP,
    long = collateralized debt position
}

\DeclareAcronym{amm}{
    short = AMM,
    long = automated market maker
}

\DeclareAcronym{dex}{
    short = DEX,
    long = decentralized exchange
}

\DeclareAcronym{ncb}{
    short = NCB,
    long = non-crypto-backed
}

\DeclareAcronym{ltv}{
    short = LTV,
    long = loan-to-value ratio
}

\DeclareAcronym{trafi}{
    short = TradFi,
    long = traditional finance
}

\DeclareAcronym{p2p}{
    short = P2P,
    long = peer-to-peer
}

\DeclareAcronym{pos}{
    short = PoS,
    long = proof-of-stake
}

\DeclareAcronym{eth}{
    short = ETH,
    long = Ether
}

\DeclareAcronym{iou}{
    short = IOU,
    long = I-owe-you
}

\DeclareAcronym{plf}{
    short = PLF,
    long = protocols for loanable funds
}

\DeclareAcronym{aum}{
    short = AUM,
    long = assets under management
}

\DeclareAcronym{abs}{
    short = ABS,
    long = asset-backed securities
}

\DeclareAcronym{cdo}{
    short = CDO,
    long = collateralized debt obligation
}

\DeclareAcronym{apr}{
    short = APR,
    long = annual percentage rate
}

\DeclareAcronym{nft}{
    short = NFT,
    long = non-fungible token
}

\DeclareAcronym{altchain}{
    short = Altchain,
    long = alternative chain
}

\DeclareAcronym{hf}{
    short = HF,
    long = health factor
}

\DeclareAcronym{cf}{
    short = CF,
    long = close factor
}

\begin{document}

\title{Piercing the Veil of TVL: DeFi Reappraised}

\author{Yichen~Luo\inst{1}\orcidlink{0009-0000-8419-5316} \and
Yebo~Feng\inst{2,3}\textsuperscript{\Letter}\orcidlink{0000-0002-7235-2377} \and
Jiahua~Xu\inst{1,3}\orcidlink{0000-0002-3993-5263} \and
Paolo~Tasca\inst{3}\orcidlink{0000-0002-5460-5940}}

\authorrunning{Y. Luo et al.}

\institute{Centre for Blockchain Technologies, University College London, London, UK\and
Nanyang Technological University, Singapore\\
\textsuperscript{\Letter} Correspondence: \email{yebo.feng@ntu.edu.sg} \and
DLT Science Foundation, London, UK
}

\newcommand\blfootnote[1]{
    \begingroup
    \renewcommand\thefootnote{}\footnote{#1}
    \addtocounter{footnote}{-1}
    \endgroup
}

\maketitle

\begin{abstract}
\Acf{tvl} is widely used to measure the size and popularity of \acf{defi}. However, \acs{tvl} can be manipulated and inflated through \enquote{double counting} activities such as wrapping and leveraging. As existing methodologies addressing double counting are inconsistent and flawed, we propose a new framework, termed \enquote{\acf{tvr}}, to assess the true underlying value of \acs{defi}. Our formal analysis reveals how \acs{defi}'s complex network spreads financial contagion via derivative tokens, increasing \acs{tvl}'s sensitivity to external shocks. To quantify double counting, we construct the \acs{defi} multiplier, which mirrors the money multiplier in \acf{trafi}. Our measurements reveal substantial double counting in \acs{defi}, finding that the gap between \acs{tvl} and \acs{tvr} reached \$139.87 billion during the peak of \acs{defi} activity on December 2, 2021, with a \acs{tvl}-to-\acs{tvr} ratio of approximately 2. We conduct sensitivity tests to evaluate the stability of \acs{tvl} compared to \acs{tvr}, demonstrating the former's significantly higher level of instability than the latter, especially during market downturns: a 25\% decline in the price of \acf{eth} leads to a \$1 billion greater decrease in \acs{tvl} compared to \acs{tvr} among leading protocols via asset value depreciation and liquidations triggered by derivative tokens. We also document that the \ac{defi} money multiplier is positively correlated with crypto market indicators and negatively correlated with macroeconomic indicators. Overall, our study suggests that \acs{tvr} is more reliable and stable than \acs{tvl}.

\end{abstract}

\section{Introduction}
\label{sec:intro}

\Acf{tvl} is one of the most widely adopted metrics for assessing both the size and popularity of \ac{defi}. Analogous to the concept of \ac{aum} in \ac{trafi}, \ac{tvl} is a similar measure of assets pooled for yield generation (see \autoref{def:tvl}) \cite{Cousaert2021,Xu2022g}. According to \href{https://defillama.com/?staking=true&pool2=true&govtokens=true&doublecounted=true&liquidstaking=true&vesting=true}{DeFiLlama}, the entire \ac{defi} \ac{tvl} stands at approximately \$250 billion as of May 24, 2025.
While blockchain systems were designed to enable automatic reconciliation and a coherent accounting whole without discrepancies~\cite{Ibanez2020c,Ibanez2025Triple-EntryKin}, \ac{defi} protocols and liquidity pools have fragmented shared ledgers into accounting \enquote{islands} where double counting thrives. 
Particularly, highly incentivized practices such as token wrapping and redepositing of borrowed tokens~\cite{Cousaert2021,Xu2022g} can trigger double counting, artificially inflating \ac{tvl} in the absence of any new capital inflows. 
Consequently, \ac{tvl} can be a deceptive metric, misleading investors to make financial decisions based on distorted valuations.\footnote{
Value inflation in the blockchain space has also been observed in other metrics, such as throughput~\cite{Perez2020}.
} 
Moreover, the culprit for double counting, the \enquote{derivative tokens} (see \autoref{def:derivative_token}), also serve as channels for financial contagion, making \ac{tvl} highly sensitive during market downturns. Unfortunately, the methods for different \ac{defi} tracing platforms to calculate \ac{tvl} are yet unstandardized, non-transparent, and often biased (see \autoref{tab:defi-tracing}), obscuring \ac{defi}'s true economic value.

\begin{table}[tb]
    \centering
    \tiny
    \caption{Survey of \ac{defi} tracing websites with a focus on protocols coverage and \ac{tvl}-related information disclosure as of February 15, 2025.}
    \begin{tabular}{lrccccccc}
\toprule

\multicolumn{1}{l}{\multirow{3}{*}{\makecell{\ac{defi} Tracing\\ Website}}} & \multicolumn{1}{l}{\makecell{Protocol\\ Coverage}} & \multicolumn{7}{c}{TVL-related Information}\\

\cmidrule(lr){2-2}\cmidrule(lr){3-9}

& \makecell{Number} & \makecell{\Ac{tvl}\\ Presented} & \makecell{Overall\\ Methodology} & \makecell{Protocol-specific\\ Methodology} & \makecell{Token Price\\ Sources} & \makecell{Constituent\\ Protocols} & Code & \makecell{Double Counting\\ Solution} \\

\hline

\href{https://docs.llama.fi/list-your-project/readme}{DeFiLlama} & 4,477 & \CIRCLE & \CIRCLE & \CIRCLE & \CIRCLE & \CIRCLE & \CIRCLE & \CIRCLE \\
\href{https://l2beat.com/faq}{L2BEAT}& N/A & \CIRCLE & \CIRCLE & \Circle & \Circle & \Circle & \CIRCLE & \Circle \\
\href{https://docs.dappradar.com/rankings/defi-rankings}{DappRadar}$^*$ & 4,588 & \halfcirc & \Circle & \Circle & \Circle & \CIRCLE & \Circle & \Circle \\
\href{https://www.stelareum.io/en/defi-tvl.html}{Stelareum} & 309 & \CIRCLE & \Circle & \Circle & \Circle & \CIRCLE & \Circle & \Circle \\
\href{https://docs.defipulse.com/methodology/tvl}{DeFiPulse} & N/A & \Circle & \CIRCLE & \Circle & \Circle & \Circle & \Circle & \CIRCLE \\
\bottomrule
\multicolumn{9}{l}{$*$: Although DappRadar tracks over 4,000 protocols, it discloses the \ac{tvl} for only around 500 ones.}\\
\multicolumn{9}{l}{\CIRCLE: Disclosure. \halfcirc: Partial Disclosure. \Circle: No disclosure.}
\end{tabular}
    \label{tab:defi-tracing}
\end{table}

The double counting problem in \ac{defi} is a crucial yet understudied topic in the literature. Many studies use \ac{tvl} for protocols valuation and risk monitoring~\cite{Kumar2022DecentralizedValuations,Maouchi2022UnderstandingNFTs,Soiman2023WhatReturns,Stepanova2021ReviewLocked} but overlook the issue of double counting within \ac{tvl}. While some existing studies document the complexity and interconnections within \ac{defi} at the token level~\cite{Saengchote2021WhereMeans.}, protocol level~\cite{Tovanich2023ContagionCompound} and sector level~\cite{Chiu2023UnderstandingModel}, most of them focus on analyzing the topological features and associated risks of \ac{defi} networks rather than their impacts on \ac{tvl}. Although a theoretical production-network model has been applied to assess the value added and service outputs across various \ac{defi} sectors on Ethereum~\cite{Chiu2023UnderstandingModel}, it fails to address double counting within individual sectors.

In this paper, we propose a novel yet effective measurement framework, termed \acf{tvr}, to address the double counting problem at the finest granularity---the token level. \ac{tvr} excludes the value of complex \ac{defi} derivatives and borrowed tokens, focusing only on the asset component that contributes directly to the underlying value of \ac{defi} that can ultimately be redeemed. By eliminating derivatives, \ac{tvr} also avoids the inclusion of financial contagion risk, making it a more stable metric than \ac{tvl}.

We reappraise the \ac{defi} system's value using the \ac{tvr} framework with data from 3,570 protocols over five years from DeFiLlama and token categories from CoinMarketCap. 
We employ the token category data fetched from CoinMarketCap to identify \enquote{plain tokens}, i.e. tokens that do not entail any underlying token. The values of these tokens are then aggregated to calculate the \ac{tvr} for the entire \ac{defi} system. Inspired by \ac{trafi} money multiplier, we introduce the \ac{defi} money multiplier, which is defined as the ratio of \ac{tvl} to \ac{tvr}. This metric quantifies the extent of double counting in \ac{defi}.
Through formalization and sensitivity tests, we compare the stability of \ac{tvl} with that of \ac{tvr}. The formalization reveals that \ac{tvl} is highly sensitive to price shocks such as \ac{eth} price decline. This sensitivity arises from the endogeneity of the derivative token price and the quantity of derivative tokens staked in \ac{plf}. We then conduct simulations to assess the stability of \ac{tvl} compared to \ac{tvr}. The simulation results align with our formalization.

We summarize our main contributions as follows:
\vspace{-3.5mm}
\begin{enumerate}\item By modeling and formalizing \ac{tvl}, we reveal the double counting mechanism and financial contagion risk under the \ac{tvl} framework. We find existing methodologies addressing double counting either inconsistent or flawed.

\item To the best of our knowledge, we are the first to introduce an enhanced measurement framework to evaluate value locked within a \ac{defi} system without double counting. Our analysis of 3,570 protocols over five years finds a substantial double counting within the DeFi system, up to \$139.87 billion with a \ac{tvl}-\ac{tvr} ratio of around 2. This contribution provides \ac{defi} users with more accurate and complete information about the value locked in \ac{defi}, which supports better decision-making within the community.

\item Our sensitivity tests reveal that \ac{tvl} is highly sensitive to market downturns compared to \ac{tvr}. A 25\% drop in \ac{eth} price leads to a significant divergence, resulting in approximately a \$1 billion greater decrease in \ac{tvl} compared to \ac{tvr} in a system of six representative \ac{defi} protocols in Ethereum.

\item We are also the first to build the \ac{defi} money multiplier based on \ac{tvr} and \ac{tvl} in parallel to the \ac{trafi} macroeconomic money multiplier to quantify the double counting. We document that the \ac{defi} money multiplier is positively correlated with crypto market indicators and negatively correlated with macroeconomic indicators.

\end{enumerate} 
\section{Key \ac{defi} Concepts }
\label{sec:key_defi_concepts}

In this section, we explain key concepts in \ac{defi}. \ac{defi} is an ecosystem of protocols operating autonomously through smart contracts, popularized by Ethereum~\cite{Hernandez2025EvolutionLiterature}. These protocols are decentralized applications inspired by traditional centralized finance systems~\cite{Xu2023SoK:Protocols}. 

\subsection{TVL}
\label{subsec:tvl}
\begin{figure}[tb]
    \begin{minipage}[tb]{0.49\linewidth}
        \centering
        \includegraphics[width=\linewidth]{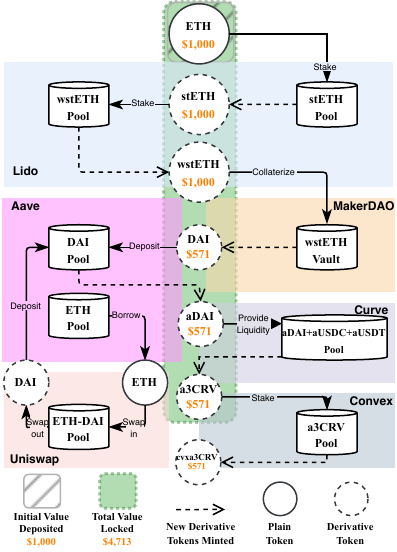}
        \caption{An example of actions a \ac{defi} user can take to maximize yield, enabled by the \ac{defi} composability.}
        \label{fig:complexity}
    \end{minipage}
    \hfill
    \begin{minipage}[tb]{0.49\linewidth}
        \centering
          \includegraphics[width=\linewidth]{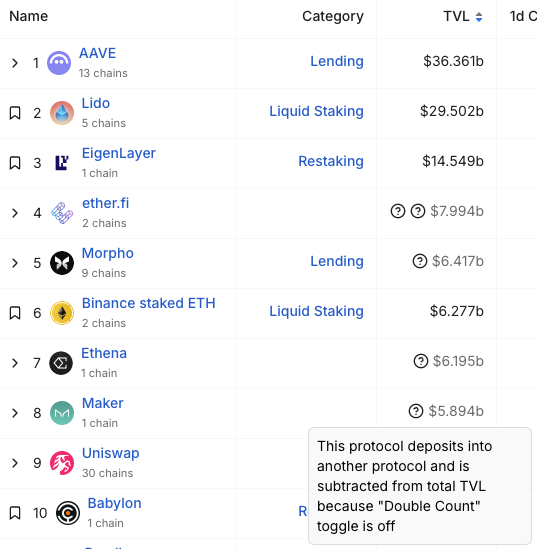}
          \caption{DeFiLlama's \ac{tvl} dashboard after deactivating the \enquote{Include in \ac{tvl}: Double Count} toggle. When a user deactivates this toggle, protocols that deposit into another protocol will be excluded from the total \ac{tvl} calculation, and their \ac{tvl} numbers will be grayed out.}
          \label{fig:double-count-removal}
    \end{minipage}
\end{figure}

Based on definitions and descriptions from existing literature~\cite{Gogol2024SoK:Risks,Xu2022b}, we define \ac{tvl} as follows: \begin{definition}[Total Value Locked]
    \label{def:tvl}
    The total value of assets staked in or held by a \ac{defi} protocol, a blockchain, or the entire \ac{defi} ecosystem at a specific moment for yield generation purposes.    
\end{definition} 
\ac{tvl} of the \ac{defi} system can be expressed as \begin{equation}
\label{eq:tvl}
{\it TVL} = \mathbf{p}^T \mathbf{Q} \mathbf{1},
\end{equation}where $\mathbf{1}$ is a column vector of ones; $\mathbf{Q} = [q_{i,j}]_{m\times n}$ denotes the matrix of staked tokens quantity across all \ac{defi} protocols, with $m$ being the number of token types and $n$ the number of \ac{defi} protocols; $\mathbf{p} = [p_{i}]_{m \times 1}$ denotes the column vector of token prices for the $m$ token types. We select Lido, MakerDAO, Aave V2, Uniswap V2, Curve, and Convex as an example system (see \autoref{fig:complexity}) to illustrate the complexity of the \ac{tvl} ecosystem. At the time of writing this paper, these protocols have the highest \ac{tvl} and are the most representative within their respective \href{https://defillama.com/categories}{category}, collectively accounting for approximately 68\% of the total \ac{tvl}. In the illustrative example in \autoref{fig:complexity}, the \ac{tvl} of these protocols totals \$4,713 --- equivalent to 4.713 times the initial ETH value deposited of \$1,000.

\ac{defi} tracing websites disclose key metrics of \ac{defi} protocols including \ac{tvl}, as shown in~\autoref{tab:defi-tracing}. DeFiLlama, a leading \ac{defi} tracing website, attempts to eliminate double counting by excluding protocols categorized under those feeding tokens into other protocols. The \enquote{Include in \ac{tvl}: Double Count} toggle allows users to filter out the \ac{tvl} of protocols that deposit into another protocol, as shown in \autoref{fig:double-count-removal}. When such protocols are excluded, the dashboard displays the chain-level DeFiLlama-adjusted \ac{tvl} (${\it TVL^{\textnormal{Adj}}}$), instead of the standard DeFiLlama \ac{tvl} that includes double counting (${\it TVL}$). While DeFiLlama makes efforts to address the issue, it may not fully eliminate double counting. Since different protocols have different degrees of double counting, simply excluding a particular category of protocols does not suffice to address the problem of double counting comprehensively.

\begin{figure}[tb]
\centering
\begin{subfigure}{.5\linewidth}
  \centering
  \includegraphics[width=0.5\linewidth]{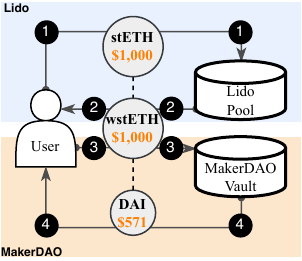}
  \caption{Wrapping in \ac{defi}.}
  \label{subfig:wrapping_defi}
\end{subfigure}%
\hfill
\begin{subfigure}{.5\linewidth}
  \centering
\includegraphics[width=0.5\linewidth]{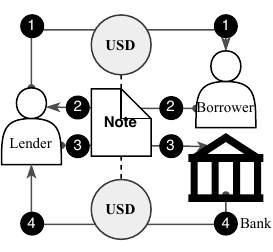}
  \caption{Rehypothecation in TradFi.}
  \label{subfig:discounting_tradfi}
\end{subfigure}
\caption{Wrapping and its corresponding \ac{trafi} analogy. The process of wrapping in \ac{defi}, as illustrated in \autoref{subfig:wrapping_defi}, mirrors the rehypothecation process in \ac{trafi}, as shown in \autoref{subfig:discounting_tradfi}. The black circle (\CIRCLE) with a white number indicates the step.}
\label{fig:double-counting}
\vspace{-5mm}
\end{figure}

\subsection{Wrapping}
\label{subsec:wrapping}

Wrapping and leveraging are two \ac{defi} mechanisms that can result in \ac{tvl} double counting. As leveraging is less prevalent and has been well documented in \cite{Perez2021Liquidations:Knife-Edge}, we provide its explanation in Appendix ~\autoref{subsec:leveraging}. Wrapping means \ac{defi} users depositing existing tokens, including tokens that have been wrapped, into smart contracts to generate new tokens. Enabled by \ac{defi} composability, users can repeatedly perform this operation to sustain their liquidity and maximize their interest \cite{Cousaert2021,Xu2022g,Xu2022d}. \ac{defi} composability refers to the ability of one \ac{defi} protocol to accept tokens generated from another protocol seamlessly, allowing \ac{defi} tokens to be chained and integrated to create new tokens and financial services. \autoref{subfig:wrapping_defi} depicts a scenario where an investor initially supplies \$1,000 worth of stETH to Lido (step \encircle{1}), which is then converted into \$1,000 worth of wstETH (step \encircle{2}). Subsequently, the investor deposits this wstETH into MakerDAO (step \encircle{3}) and issues up to \$571 worth of DAI (step \encircle{4})\footnote{The calculation of DAI amount is based on the \ac{ltv} of the wstETH low fee vault at the time of this paper and the collateral value.}. The wrapping in \ac{defi} is similar to the rehypothecation process in \ac{trafi}, illustrated in \autoref{subfig:discounting_tradfi}. This works as follows. In rehypothecation, lenders who provide loans to borrowers (step \encircle{1}) against a promissory note (step \encircle{2}), pledge the promissory note (step \encircle{3}) and borrow money from the bank (step \encircle{4})~\cite{Andolfatto2017RehypothecationLiquidity}.

To illustrate the \ac{tvl} double counting during wrapping, we use a balance-sheet approach to consolidate double-entry bookkeeping (see e.g. \cite{Richardson2020e}) and describe its financial condition for each protocol. Appendix~\autoref{sec:defi_bookkeeping} provides a detailed list of bookkeeping entries for common transactions in \ac{defi} protocols. The aggregate value locked can be regarded as a significant element on the asset side of a \ac{defi} protocol's balance sheet~\cite{Xu2025Auto.gov:DeFi}. In the context of a \ac{defi} system, we apply the principles of consolidated balance sheets to depict its financial status on an aggregated basis. This type of balance sheet represents the combined financial position of a group, presenting the assets, liabilities, and net position of both the parent company and its subsidiaries as those of a unified economic entity. By employing the principle of non-duplication used in consolidated balance sheet accounting (whereby, accounting entries that are recorded as assets in one company and as liabilities in another are eliminated, before aggregating all remaining items)~\cite{Simmonds1986ConsolidatedSubsidiary}, we can effectively eliminate instances of double counting within a \ac{defi} system. \autoref{tab:consolidated-balance-sheet} shows the balance sheets of Lido and MakerDAO, and the consolidated balance sheet of the \ac{defi} system consists of Lido and MakerDAO. Immediately after step \encircle{2}, the value of the DeFi system is \$1,000. However, if we deposit the receipt token wstETH from Lido into MakerDAO to issue another receipt token DAI, the \ac{tvl} will be \$2,000 under the traditional TVL measurement. The balance sheets are expanded and the \ac{tvl} is double-counted due to the existence of wstETH. In the consolidated balance sheet, the \ac{tvl} is adjusted to \$1,000 after eliminating the value associated with the wstETH.

\begin{table}[tb]
\centering
\tiny
\caption{Protocol-perspective balance sheets of the wrapping scenario. We \textbf{boldface} the components included in the \ac{tvl} calculation.}
\label{tab:consolidated-balance-sheet}
\begin{subtable}{0.32\linewidth}
\centering
\caption{Lido.}
\label{subtab:balance-sheet-lido}
\begin{tabular}{@{}p{0.63\linewidth}>{\raggedleft\arraybackslash}p{0.14\linewidth}>{\raggedleft\arraybackslash}p{0.17\linewidth}@{}}
\toprule
\textbf{Immediately After} & \encircle{2} & \encircle{4} \\
\midrule
\textit{Assets} & \$ & \$ \\
Value Locked - stETH & \textbf{1,000} & \textbf{1,000} \\
&&\\
\midrule
\textbf{Total Assets} & 1,000 & 1,000 \\
\bottomrule\bottomrule
\textit{Liabilities} & \$ & \$ \\
Payables - wstETH & 1,000 & 1,000 \\
&&\\
\midrule
\textbf{Total Liabilities} & 1,000 & 1,000 \\
\bottomrule\bottomrule
\end{tabular}
\end{subtable}%
\hfill
\begin{subtable}{0.32\linewidth}
\centering
\caption{MakerDAO.}
\label{subtab:balance-sheet-makerdao}
\begin{tabular}{@{}p{0.67\linewidth}>{\raggedleft\arraybackslash}p{0.14\linewidth}>{\raggedleft\arraybackslash}p{0.17\linewidth}@{}}
\toprule
\textbf{Immediately After} & \encircle{2} & \encircle{4} \\
\midrule
\textit{Assets} & \$ & \$ \\
Value Locked - wstETH & - & \textbf{1,000} \\
Receivables - DAI & - & 571 \\
\midrule
\textbf{Total Assets} & - & 1,571 \\
\bottomrule\bottomrule
\textit{Liabilities} & \$ & \$ \\
Payables - wstETH & - & 1,000 \\
New Money - DAI & - & 571 \\
\midrule
\textbf{Total Liabilities} & - & 1,571 \\
\bottomrule\bottomrule
\end{tabular}
\end{subtable}%
\hfill
\begin{subtable}{0.33\linewidth}
\centering
\caption{Consolidated.}
\label{subtab:consolidated-balance-sheet}
\begin{tabular}{@{}p{0.53\linewidth}>{\raggedleft\arraybackslash}p{0.14\linewidth}>{\raggedleft\arraybackslash}p{0.17\linewidth}@{}}
\toprule
\textbf{Immediately After} & \encircle{2} & \encircle{4} \\
\midrule
\textit{Assets} & \$ & \$ \\
Value Locked & \textbf{1,000} & \textbf{1,000} \\
Receivables & - & 571 \\
\midrule
\textbf{Total Assets} & 1,000 & 1,571 \\
\bottomrule\bottomrule
\textit{Liabilities} & \$ & \$ \\
Payables & 1,000 & 1,571 \\
&&\\
\midrule
\textbf{Total Liabilities} & 1,000 & 1,571 \\
\bottomrule\bottomrule
\end{tabular}
\end{subtable}
\vspace{-5mm}
\end{table}

\section{Enhanced Measurement Framework}

\label{sec:enhanced_measurement_framework}

\label{sec:defi_token_classification}
In this section, we formalize the double counting problem, identify the instability of \ac{tvl}, and introduce our enhanced measurement framework \ac{tvr}. 

\subsection{Double Counting Problem}
We initially classify \ac{defi} tokens into plain tokens and derivative tokens.  We provide the following definitions for the plain and underlying tokens:\begin{definition}[Plain Token]
    \label{def:plain_token}
    A \ac{defi} token without any underlying token.
\end{definition} \begin{definition}[Derivative Token]
    \label{def:derivative_token}
    A \ac{defi} receipt token, also known as an \ac{iou} token or a depositary receipt token, that is generated in a specified ratio by depositing some underlying tokens into the smart contract of a protocol.
\end{definition}In the example used in \autoref{fig:complexity}, \ac{eth} is considered as a plain token because it is initially deposited into the system without having an underlying token. In contrast, other tokens (stETH, wstETH, DAI, aDAI, a3CRV, and cvxa3CRV) are considered derivative tokens since they have underlying tokens. Let $\bm{\tau} = [\tau_{i}]_{m\times 1}$, where $m$ is the number of unique tokens and $\tau_{i} = 1$ if token $i$ is a plain token and 0 otherwise. By decomposing the \ac{tvl} value, as defined in \autoref{eq:tvl}, into the plain token value and the derivative token value, we can further express the \ac{tvl} as ${\it TVL} = (\mathbf{p} \odot \bm{\tau})^T \mathbf{Q} \mathbf{1} + [\mathbf{p} \odot (\mathbf{1} - \bm{\tau})]^T \mathbf{Q} \mathbf{1} $, where $\odot$ denotes the element-wise product. Since the value of derivative tokens can be easily created and inflated through wrapping without injecting any capital into the \ac{defi} system according to \autoref{subsec:wrapping}, there is an urgent need to design a new framework that excludes this inflated value. The new framework should ensure that the underlying value, which cannot be easily manipulated, is accurately reflected. 

\subsection{Instability of \ac{tvl}}
\label{sec:endo_price}
In addition to the inflation issue, derivative tokens can act as channels for the spread of financial contagion, making the \ac{tvl} highly sensitive to market downturns. The prices of derivative tokens are endogenously determined by the prices and quantities of their underlying tokens. The pricing mechanism can be described as follows: \begin{enumerate*}[label=(\roman*)] \item If the derivative token is a stablecoin generated from a \ac{cdp} and the aggregate value of its underlying tokens meets or exceeds the value of the stablecoin, the token's price is pegged to its predetermined fiat currency. This peg is maintained through an overcollateralization mechanism, as discussed in Appendix~\autoref{subsec:liquidation}. \item Otherwise, the token price is determined by the ratio of the underlying tokens' total value to the derivative token's circulating supply. \end{enumerate*} In both cases, a short-term fluctuation term $\epsilon_d$, should be included to account for temporary price variations.

We write the derivative token price as \begin{equation}
    \label{eq:endo_price} 
    p_{d}(\textbf{p}_{u}, \textbf{q}_{u}) = \begin{cases} 
      c_{d} & \text{if $d$ is a CDP stablecoin and } \Gamma \geq c_d\\
      \Gamma & \text{otherwise}
    \end{cases} \qquad + \epsilon_{d},
\end{equation}where $\Gamma = \frac{\mathbf{p}_u^T \mathbf{q}_u}{q_d}$. $p_{d}$ and $q_{d}$ are the price and circulating supply of derivative token $d$. $\mathbf{p}_u = \left[p_i\right]_{m \times 1}$ and $\mathbf{q}_u = \left[q_i\right]_{m \times 1}$ are the vector of $d$'s underlying token prices and quantities, respectively. $c_d$ is the theoretical pegging price of a \ac{cdp} stablecoin in USD. The short-term fluctuation $\epsilon_{d}$ is exogenous, associated with the token's supply and demand dynamics as well as the liquidity. For example, the price of stETH deviated from its reference point temporarily in 2022 due to selling pressure from \href{https://medium.com/huobi-research/steth-depegging-what-are-the-consequences-20b4b7327b0c}{Celsius and market illiquidity}. Appendix~\autoref{derivative_token_pegging_mechanism} explains in detail the derivative token pegging mechanism that supports \autoref{eq:endo_price}. For \ac{cdp} stablecoins, the deviation of their price, i.e. \enquote{depegging}, from the predetermined reference point due to the undercollateralization, is denoted as $\Gamma < c_d$.

\ac{defi} composability allows the underlying token of a derivative token to serve as the derivative token of another token, as illustrated in \autoref{fig:complexity} and \autoref{subfig:wrapping_defi} (e.g. wstETH is a derivative token of stETH, which itself is a derivative token of \ac{eth}). We can derive the derivative token price function in terms of its ultimate underlying plain token prices and quantities: $p_{d}(\mathbf{p}_u,\mathbf{q}_u) = \left[p_{d_1} \circ p_{d_2} \ldots \circ p_{d_j}\right](\mathbf{p}_u,\mathbf{q}_u)$, where $\mathbf{p}_u = \left[p_i\right]_{v \times 1}$ is the vector of $d$'s ultimate underlying token prices via the recursion of \autoref{eq:endo_price}. $\circ$ is the function composition operator. $[p_{d_1} \circ p_{d_2}  \ldots \circ p_{d_j}]$ means we recurse \autoref{eq:endo_price} multiple times until we find the ultimate underlying plain tokens (e.g. ETH as the ultimate plain token of wstETH).

\label{sec:endo_quantity}

Tokens staked in a \ac{plf}, including \acp{cdp} such as MakerDAO or lending protocols such as Aave, have a token quantity affected by its token price due to the liquidation mechanism \cite{Arora2024SecPLF:Attacks,Xu2025Auto.gov:DeFi,Xu2022d}. Detailed definitions of \ac{plf} and its liquidation mechanism are provided in Appendix~\autoref{subsec:liquidation}. According to Appendix~\autoref{subsec:liquidation}, a change in collateral $j$'s price $p_{j,t} \rightarrow p_{j,t+1}$ will lead to the change of the account $i$'s health factor $h_{i,t+1}(p_{j,t+1})$, a ratio between liquidation threshold-adjusted collateral value to debt value, and the liquidation profit $\Pi_{i,t+1}(p_{j,t+1})$, leading to different scenarios. In liquidation, we should consider the quantity of both collateral tokens and repaid tokens since the liquidator not only withdraws collaterals but also injects liquidity into the protocol via the repayment.

When $h_{i,t+1}(p_{j,t+1}) \geq 1$, the account is deemed safe and the quantity of collateral $j$ in the account remains unchanged, represented by $q_{i,j,t+1}$. When $h_{i,t+1}(p_{j,t+1}) < 1$ and the liquidation profit $\Pi_{i,t+1}(p_{j,t+1}) \leq 0$, the liquidation is considered unprofitable for liquidators, rendering the liquidation unviable and the quantity of collateral in the account also unchanged, represented by $q_{i,j,t+1}$. 

When $h_{i,t+1}(p_{j,t+1}) < 1$, the user may face liquidation, where the smart contract transfers and sells varying proportions of collateral to maintain the solvency of \ac{plf}. Additionally, when the liquidation profit $\Pi_{i,t+1}(p_{j,t+1}) > 0$, the total collateral value is sufficient to cover the total debt value. In this scenario, the liquidation is deemed profitable for liquidators, leading to a successful liquidation. In a liquidation, the token quantity obeys the following law of motion when $t \rightarrow t+1$:\begin{equation}
    \label{eq:endo_quantity}
     q_{i,j,t+1}(p_{i,j,t+1}) = \begin{cases} 
     q_{i,j,t} + \Delta_{i,j,t+1} & \text{if } h_{i,t+1} < 1 \text{ and } \Pi_{i,t+1} > 0\\
     q_{i,j,t} & \text{otherwise}
    \end{cases}.
\end{equation}
$\Delta$ and $\Pi$ depend on the type of \ac{plf} and tokens as shown in \autoref{tab:Delta}, as explained in Appendix~\autoref{derivation_delta}.


Given the endogeneity mentioned above, we can then further split the \ac{tvl} into the following four categories: the value of plain tokens staked in non-\acp{plf}, plain tokens staked in \acp{plf}, derivative tokens staked in non-\acp{plf}, and derivative tokens staked in \acp{plf}:%
\begin{equation}
\scriptsize
    \label{eq:category_tvl}
    {\it TVL} = \underbrace{(\mathbf{p} \odot \bm{\tau})^T \mathbf{Q} (\mathbf{1} - \bm{\omega})}_{\substack{\text{\scriptsize plain tokens}\\\text{\scriptsize staked in non-\acp{plf}}}} + \underbrace{(\mathbf{p} \odot \bm{\tau})^T \mathbf{Q} \bm{\omega}}_{\substack{\text{\scriptsize plain tokens}\\\text{\scriptsize staked in \acp{plf}}}} + \underbrace{[\mathbf{p} \odot (\bm{1} - \bm{\tau})]^T\mathbf{Q} (\mathbf{1} - \bm{\omega}) }_{\substack{\text{\scriptsize derivative tokens}\\\text{\scriptsize staked in non-\acp{plf}}}} + \underbrace{[\mathbf{p} \odot (1 - \bm{\tau})]^T\mathbf{Q} \bm{\omega}}_{\substack{\text{\scriptsize derivative tokens}\\\text{\scriptsize staked in \acp{plf}}}},
\end{equation}%
where $\bm{\omega} = [\omega_{i}]_{n \times 1}$, $\omega_{i} = 1$ if protocol $i$ is a \ac{plf} and 0 otherwise. The derivative token price depends on its underlying token's price $\mathbf{p}_u$ and quantity (see \autoref{eq:endo_price}). In addition, the tokens quantity staked within a \ac{plf} is affected by their own price (see \autoref{eq:endo_quantity}). Therefore, the \ac{tvl} ultimately depends on the prices and quantities of the underlying tokens. Price and quantity shocks to the underlying tokens can lead to a decline in token value, trigger liquidations, and cause depegging for derivative tokens due to the endogenous relationship between derivative and underlying tokens. Derivative tokens amplify the impact of such shocks on the \ac{tvl}, making the \ac{tvl} highly sensitive to changes in the prices of plain tokens. Consequently, the existence of derivative tokens not only inflates the \ac{tvl} but also serves as the channel for the spread of decentralized financial contagion, making the \ac{tvl} unstable. 

\label{subsubsec:metrics-inflation-and-decentralized-financial-contagion}

\begin{table}[tb]
    \caption{$\Delta$ and $\Pi$ in different scenarios, where $V_{\textnormal{liq}} = \min\{\frac{V_c}{1+b}, \delta \cdot V_d\}$ represents the maximum amount of debt that a liquidator can repay at a single liquidation in a lending protocol. $b$ represents the liquidation bonus. $\delta$ denotes the close factor. $V_c = \mathbf{c}^T \mathbf{p}_c$ and $V_d = \mathbf{d}^T \mathbf{p}_d$, where $\mathbf{c}$ and $\mathbf{d}$ are vectors of collateral and debt token quantities, represent the total collateral value and total debt value of the position, respectively, as mentioned in Appendix~\autoref{subsec:liquidation}. $gasFees$ denotes the gas costs of liquidation.}
    \centering
    \scriptsize
    \begin{subtable}{0.48\linewidth}
        \centering
        \caption{Quantity increase $\Delta_{t+1}$ for token in a CDP or a lending protocol.}
        \begin{tabular}{c|c|c}
\hline
& \textbf{\makecell{$\bm{\Delta}$ of\\Repaid Token}} & \textbf{\makecell{$\bm{\Delta}$ of\\Collateral Token}}\\
\hline
\textbf{\makecell{CDP}} & $0$ & $- q$ \\
\hline
\textbf{\makecell{Lending\\Protocol}} &  $\frac{V_{\textnormal{liq}}\cdot q}{V_d}$ & $ - \frac{(1+b)V_{\textnormal{liq}}\cdot q}{V_{c}}$\\
\hline
\end{tabular} 
        \label{tab:Delta}
    \end{subtable}%
    \hfill
    \begin{subtable}{0.48\linewidth}
        \centering
        \caption{Liquidation profit $\Pi_{t+1}$ in a CDP or a lending protocol.}
        \begin{tabular}{c|c}
\hline
 & \multirow{2}{*}{$\bm{\Pi}$} \\
 & \\
\hline
\textbf{\makecell{CDP}} & $V_c - V_d - {\it gasFees}$  \\
\hline
\textbf{\makecell{Lending\\Protocol}} &  $V_{\textnormal{liq}} \cdot b  - {\it gasFees}$\\
\hline
\end{tabular}
        \label{tab:Profit}
    \end{subtable}
    
\end{table}

\subsection{Total Value Redeemable (TVR)}
\label{subsec:tvr}
To address the double counting problem, we introduce the metric \ac{tvr}. 

\begin{definition}[Total Value Redeemable]
    \label{def:tvr}
    Token value that can be ultimately redeemed from a \ac{defi} protocol or a \ac{defi} ecosystem.
\end{definition}

We can express the \ac{tvr} of the entire \ac{defi} ecosystem as the sum of the total value of plain tokens including governance tokens, native tokens, and \ac{ncb} stablecoins held by smart contracts in the \ac{defi} ecosystem:\begin{equation}
\label{eq:tvr}
{\it TVR} = (\mathbf{p} \odot \bm{\tau})^T \mathbf{Q}' \mathbf{1} = \underbrace{(\mathbf{p} \odot \bm{\tau})^T \mathbf{Q}' (\mathbf{1} - \bm{\omega})}_{\substack{\text{\scriptsize plain tokens held by}\\ \text{\scriptsize smart contracts in non-\acp{plf}}}} + \underbrace{(\mathbf{p} \odot \bm{\tau})^T \mathbf{Q}' \bm{\omega}}_{\substack{\text{\scriptsize plain tokens held by}\\\text{\scriptsize smart contracts in \acp{plf}}}},
\end{equation}where $\mathbf{Q}' = [q'_{i,j}]_{m\times n}$, with $m$ representing the number of token types and $n$ is the number of \ac{defi} protocols, denotes the matrix of tokens quantity held by smart contracts across all \ac{defi} protocols. Compared to \ac{tvl}, \ac{tvr} excludes the value of derivative and borrowed tokens, considering only the value of plain tokens held by smart contracts to address the double counting problem. The exclusion of inflated values also decreases the complexity of the interplay within the \ac{defi} system, mitigating the high sensitivity of the metric concerning the ultimate underlying plain tokens. We also introduce the protocol-level \ac{tvr} to address the intra-protocol double counting, as discussed in Appendix~\autoref{ind_tvl}. 

DeFiLlama provides \ac{tvl} adjusted for double counting of blockchains. Additionally, it aggregates chain-level \ac{tvl} to compute the adjusted \ac{tvl} for the entire \ac{defi} ecosystem. However, it does not offer adjusted \ac{tvl} for specific protocols. Compared to DeFiLlama's adjusted \ac{tvl}, \ac{tvr} eliminates double counting with finer granularity by selectively including or excluding tokens during the calculation, resulting in significantly higher accuracy. We provide a detailed comparison in calculation methods between \ac{defi} space ecosystem-wide \ac{tvr} and DeFiLlama-adjusted \ac{tvl} in Appendix~\autoref{compare}.

To examine the stability of \ac{tvl} and \ac{tvr}, we perform comparative sensitivity analyses on the changes in \ac{tvl} ($\Delta {\it TVL}_{t+1} = {\it TVL}_{t+1} - {\it TVL}_{t}$) and \ac{tvr} ($\Delta {\it TVR}_{t+1} = {\it TVR}_{t+1} - {\it TVR}_{t}$) in response to shocks in the price of plain tokens. These tests are conducted using six representative protocols, as selected in \autoref{subsec:tvl}. For the plain token price shock, we use the decline in \ac{eth} price as the independent variable because \ac{eth} is the native token of Ethereum and is widely used across the \ac{defi} platform. Subsequently, we update the token price vector $\mathbf{p}$ from \autoref{eq:endo_price}, quantity matrix $\mathbf{Q}$ from \autoref{eq:endo_quantity}, and $\mathbf{Q}'$ from \autoref{eq:tvr}. Finally, we calculate $\Delta {\it TVL}_{t+1}$ and $\Delta {\it TVR}_{t+1}$.

\section{Empirical Analyses}
\label{sec:results}

This section details the data used for measurements and presents empirical results under both the traditional \ac{tvl} framework and our proposed \ac{tvr} framework. We also introduce the \ac{defi} money multiplier to quantify double counting in \ac{defi} and provide measurement results for individual altchains.

\subsection{Data}
\label{subsec:data_collection}

We fetch the \ac{tvl} data about \ac{defi} protocols broken down by token and adjust \ac{tvl} (${\it TVL}^{\textnormal{Adj}}$) from January 1, 2021 to March 1, 2024 using DeFiLlama API. DeFiLlama offers the most comprehensive universe of DeFi protocols of all blockchains compared to all other DeFi-tracing websites, as discussed in~\autoref{tab:defi-tracing}. ${\it TVL}^{\textnormal{Adj}}$ is DeFiLlama's improved metric aimed at mitigating the double counting problem and is flawed as discussed in \autoref{subsec:tvl}. We then break down the \ac{tvl} of each protocol to obtain the unadjusted \ac{tvl} per protocol per day (${\it TVL}_{i}$). Additionally, we retrieve token categorization lists for native tokens (\href{https://coinmarketcap.com/view/layer-1/}{layer-one} and \href{https://coinmarketcap.com/view/layer-2/}{layer-two}) and \href{https://coinmarketcap.com/view/governance/}{governance tokens} from CoinMarketCap, and obtain stablecoin classifications from \href{https://defillama.com/stablecoins}{DeFiLlama}. These lists are then used as filters to extract plain tokens from the \ac{tvl} breakdown data provided by DeFiLlama to calculate the \ac{tvr} (${\it TVR}$). We also retrieve the blockchain states from an Ethereum archive node for three key dates: December 2, 2021, marking the peak of DeFiLlama-unadjusted \ac{tvl}; May 9, 2022, denoting the end of the Luna collapse; and November 8, 2022, representing the end of the FTX collapse. In Appendix~\autoref{automation}, we explore methods to automate this process and eliminate reliance on third-party data.

For the risk analysis, we retrieve the data by crawling blockchain states (e.g. MakerDAO vaults data) and blockchain events (e.g. Aave deposit events) from an Ethereum archive node. Our sample of risk analysis constitutes six leading \ac{defi} protocols with the highest \ac{tvl} within each respective \ac{defi} protocol category, as shown in \autoref{fig:complexity}. Appendix~\autoref{sec:params} reports the statistics of accounts in sensitivity tests in MakerDAO and Aave on three representative dates. 

\begin{figure}[t]
    \centering
    \includegraphics[width=0.8\linewidth]{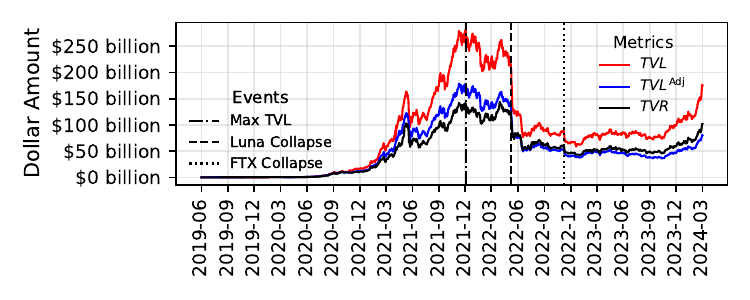}
    \vspace{-5mm}
    \caption{\ac{tvl} and \ac{tvr} over time, where the red, blue, and black lines represent the ${\it TVL}$, ${\it TVL}^{\textnormal{Adj}}$, and ${\it TVR}$.}
    \label{fig:tvl_Total}
\end{figure}

\subsection{\Ac{tvl}, Adjusted \ac{tvl}, and \ac{tvr}}
\label{subsec:tvl_tvr}

Based on the enhanced measurement framework, we build \ac{tvr} from DeFiLlama \ac{tvl} breakdown data. Our framework calculates \ac{defi} space ecosystem-wide \ac{tvr} by summing the value of all \textbf{eligible tokens} as described in \autoref{subsec:tvr}, specifically plain tokens. In contrast, DeFiLlama-adjusted \ac{tvl} is calculated by first aggregating the \ac{tvl} of all \textbf{eligible protocols} and then summing the \ac{tvl} across all blockchains, where protocol eligibility is arbitrarily determined by DeFiLlama whose validity we challenge. For instance, although MakerDAO holds both plain tokens (e.g. \ac{eth}) that directly contribute to DeFi's underlying value and derivative tokens (e.g. wstETH) that should be excluded, its entire \ac{tvl} is excluded from the DeFiLlama-adjusted \ac{tvl}, as illustrated in \autoref{fig:double-count-removal}. Appendix~\autoref{compare} conducts a detailed comparison in calculation methods between system-wide \ac{tvr} and DeFiLlama-adjusted \ac{tvl}. \autoref{fig:tvl_Total} shows the DeFiLlama unadjusted \ac{tvl} (${\it TVL}$), DeFiLlama-adjusted \ac{tvl} (${\it TVL}^{\textnormal{Adj}}$), and \ac{tvr} (${\it TVR}$) for the entire blockchain ecosystem over time. Our empirical measurement reveals the level of double counting within the DeFi ecosystem, with \acs{tvl}-\acs{tvr} discrepancies reaching up to \$139.87 billion, and a \acs{tvl}-\acs{tvr} ratio of around 2 when the unadjusted \acs{tvl} reached its maximum value. Moreover, there is a divergence between DeFiLlama-adjusted \ac{tvl} and the \ac{tvr} due to differences in methodology. In June 2022, the \ac{tvr} exceeds DeFiLlama-adjusted \ac{tvl} because the token value deposited of removed protocols under DeFiLlama's methodology is\begin{multicols}{2}
\begin{figure}[H]
    \centering
    \begin{subfigure}{\linewidth}
        \centering
        \includegraphics[width=\linewidth]{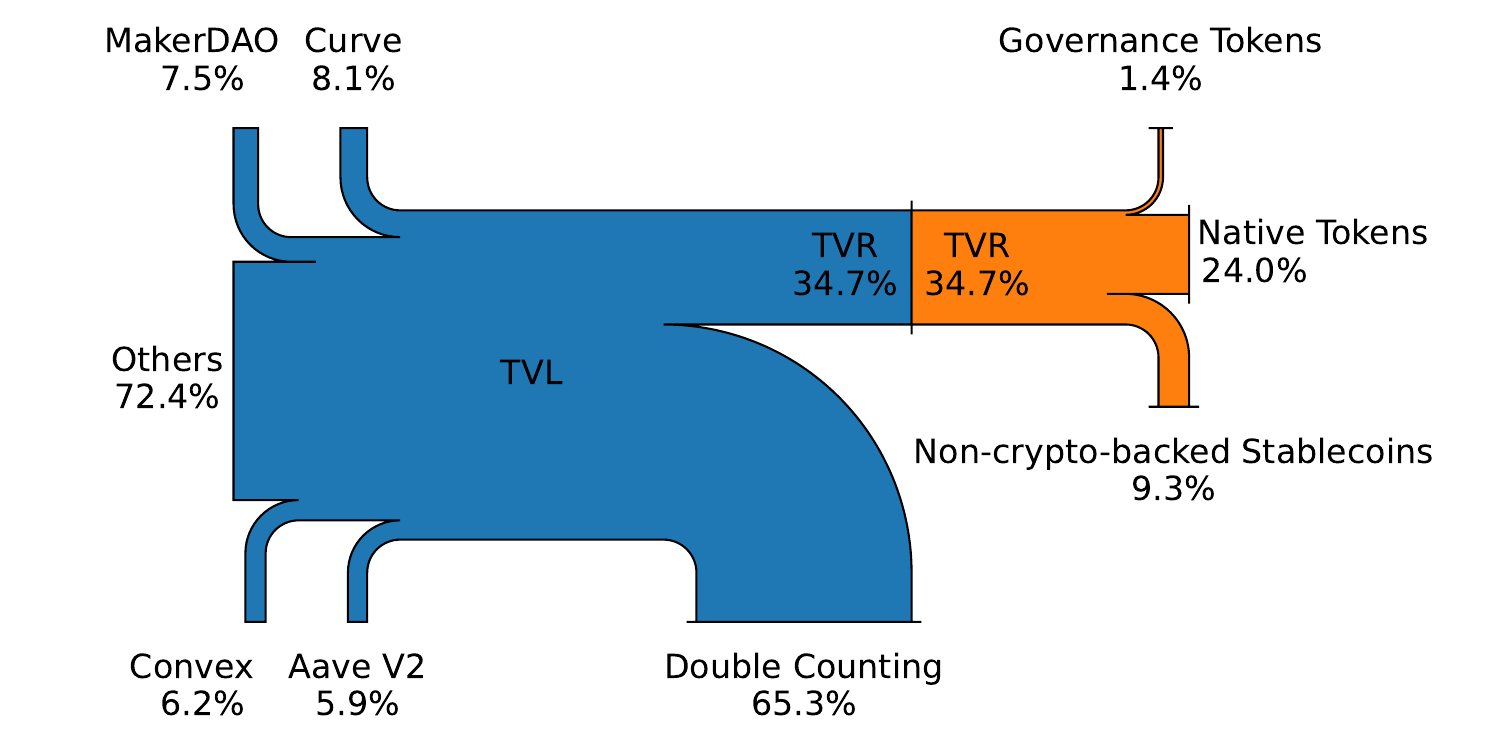}
        \vspace{-7mm}
        \caption{Max \ac{tvl}.}
        \label{subfig:tvl_decompose_sankey_max_tvl}
    \end{subfigure}%
    
    \begin{subfigure}{\linewidth}
        \centering
        \includegraphics[width=\linewidth]{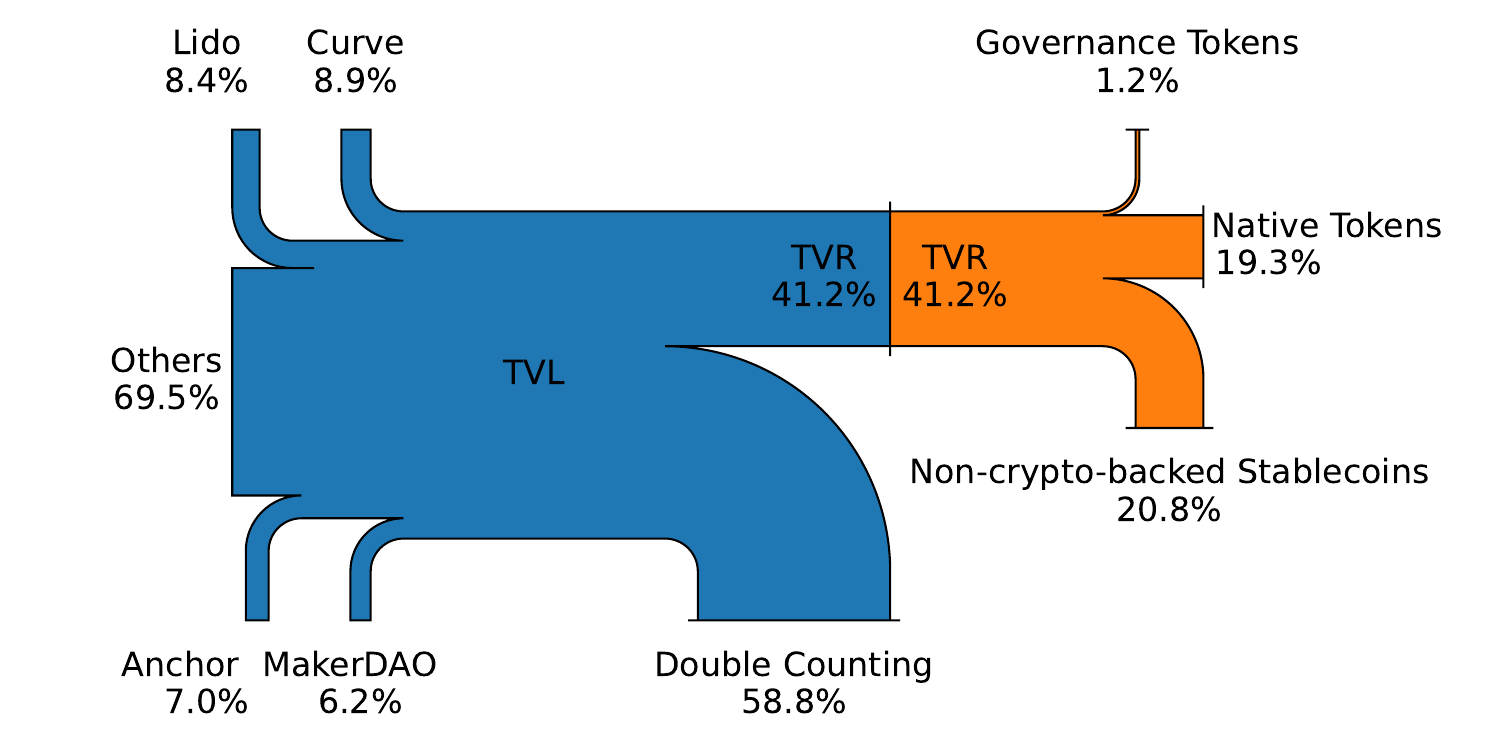}
        \vspace{-7mm}
        \caption{Luna Collapse.}
        \label{subfig:tvl_decompose_sankey_luna_collapse}
    \end{subfigure}

    \begin{subfigure}{\linewidth}
        \centering
        \includegraphics[width=\linewidth]{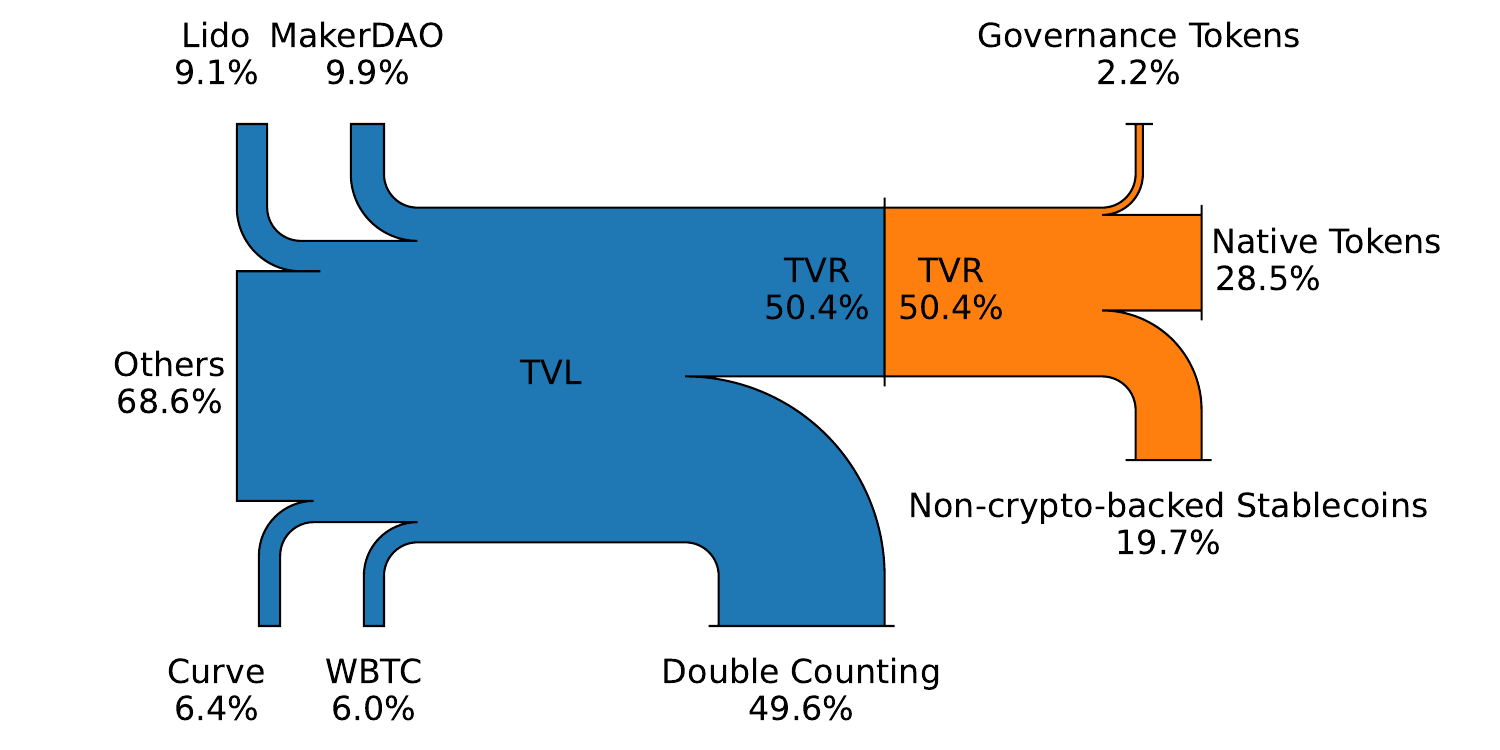}
        \vspace{-7mm}
        \caption{FTX collapse.}
        \label{subfig:tvl_decompose_sankey_ftx_collapse}
    \end{subfigure}
    \vspace{-7mm}
    \caption{Decomposition of \ac{tvl} of the entire \ac{defi} system. We identify four protocols with the highest \ac{tvl} and group the remaining protocols under the category of \enquote{Others}. The band width represents the dollar value of tokens. The blue band represents the \ac{tvl} value, while the orange band represents the \ac{tvr} value.}
    \label{fig:tvl_decompose_sankey}
\end{figure}
\begin{figure}[H]
\centering
    \begin{subfigure}{0.94\linewidth}
      \centering
      \includegraphics[width=\linewidth]{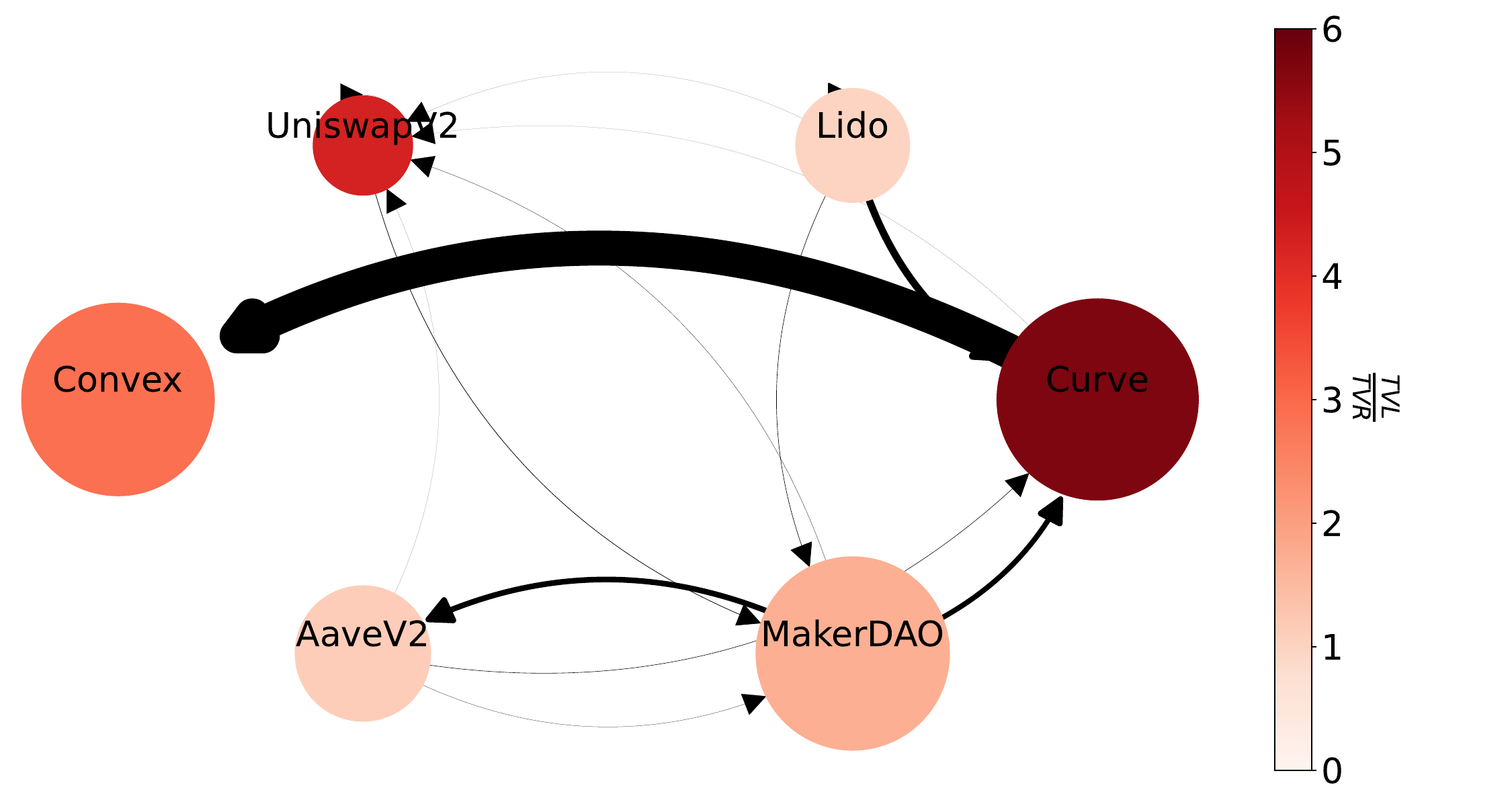}
      \vspace{-7mm}
      \caption{Max \ac{tvl}.}
      \label{subfig:network_max_tvl}
    \end{subfigure}%
    
    \begin{subfigure}{0.94\linewidth}
      \centering
      \includegraphics[width=\linewidth]{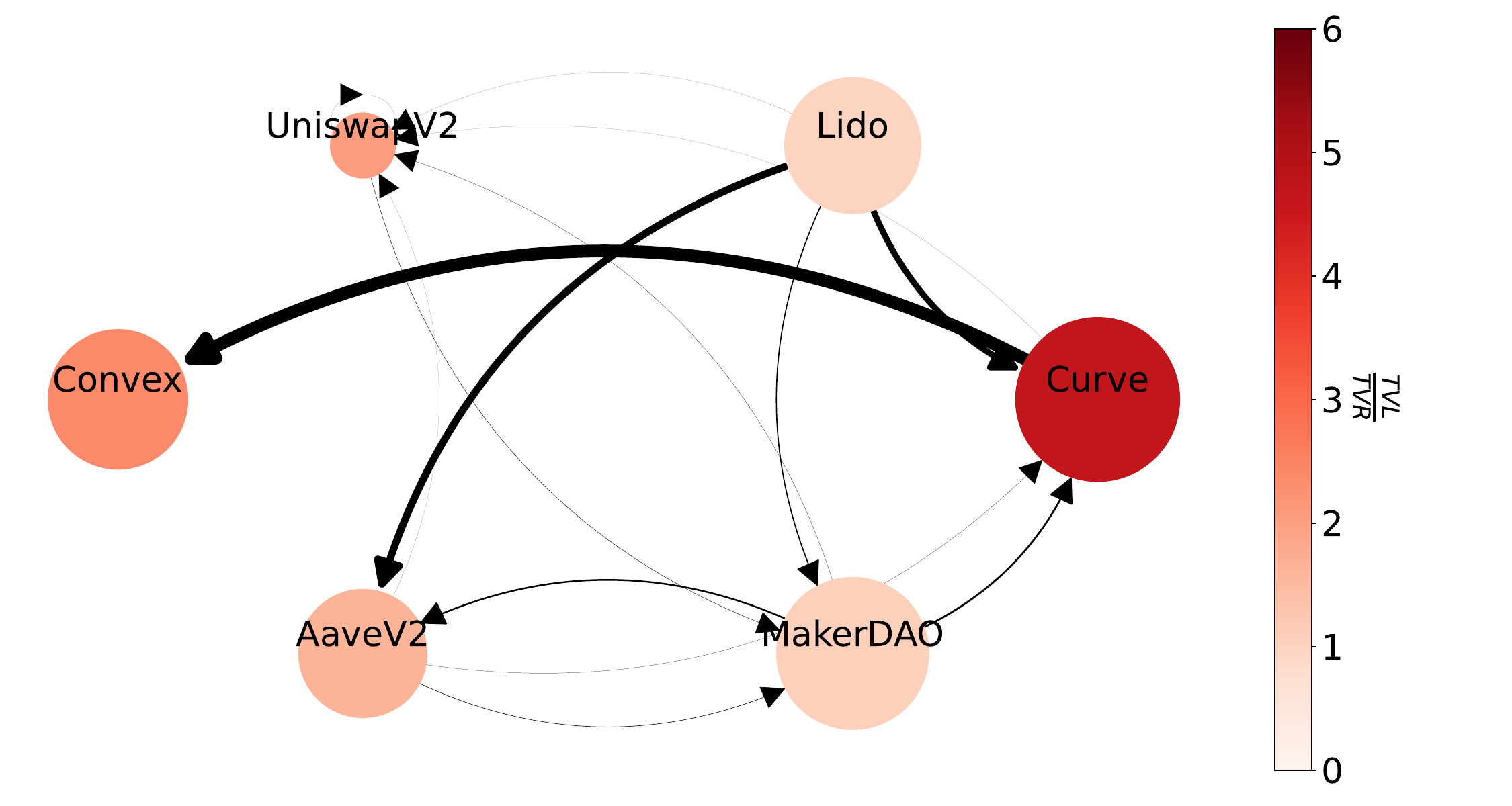}
      \vspace{-7mm}
      \caption{Luna Collapse.}
      \label{subfig:network_luna_collapse}
    \end{subfigure}%
    
    \begin{subfigure}{0.94\linewidth}
      \centering
      \includegraphics[width=\linewidth]{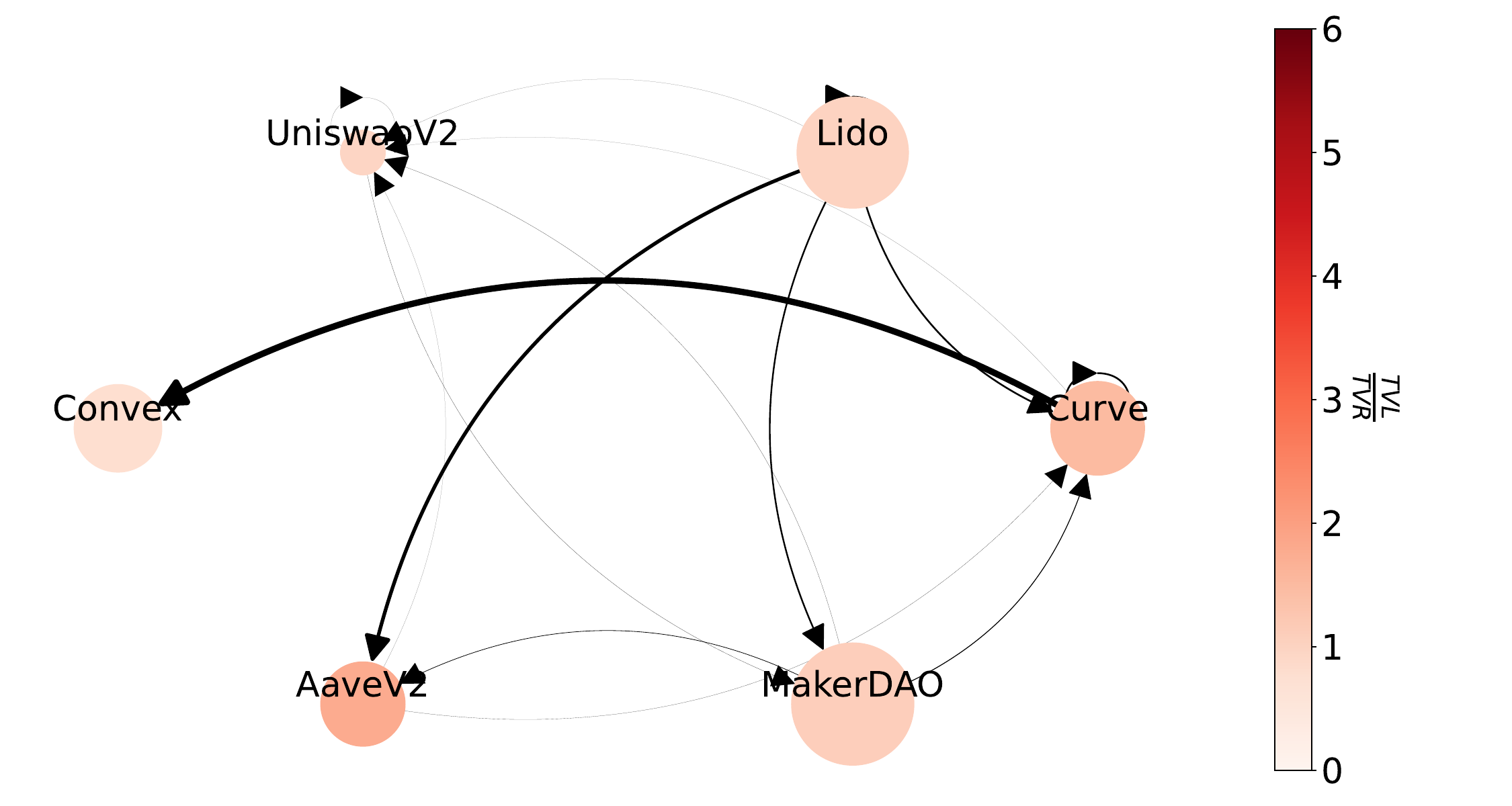}
      \vspace{-7mm}
      \caption{FTX Collapse.}
      \label{subfig:network_ftx_collapse}
    \end{subfigure}
\vspace{-3mm}
\caption{Token wrapping network of six representative protocols. Node size corresponds to the \ac{tvl}, edge width represents the dollar amount of tokens generated from the source protocol and staked in the target protocol, and node color reflects the ratio between \ac{tvl} and \ac{tvr}. A darker color indicates a higher level of double counting.}
\label{fig:protocol_network}  
\end{figure}
\end{multicols}\noindent lower than the actual value that needs to be removed within the \ac{tvr} framework. Conversely, after June 2022, the \ac{tvr} falls below DeFiLlama's adjusted \ac{tvl} because the token value deposited of protocols removed by DeFiLlama is higher than the actual value that needs to be removed within the \ac{tvr} framework. This discrepancy highlights the inaccuracies in DeFiLlama's methodology, which we document in \autoref{subsec:tvr}.

However, all three metrics show a similar trend, with a surge during the DeFi summer due to increased investor activity and sharp declines following the Luna collapse and the FTX collapse. \autoref{fig:tvl_decompose_sankey} illustrates the decomposition of \ac{tvl} in the \ac{defi} system on three key dates. The dominance of the top four protocols increases following the collapse of the Luna and FTX, while the double-counting proportion decreases after these events. The proportion of governance tokens in \ac{tvr} remains small. The proportion of native tokens decreases after the Luna collapse but increases following the FTX collapse. Conversely, the proportion of \ac{ncb} stablecoins rises after the Luna collapse but declines after the FTX collapse. To gain a broader understanding of the double-counting issue, we also analyze two niche \acp{altchain} in Appendix~\autoref{altchain_analyses}.
\begin{figure}[tb]
    \centering
    \includegraphics[width=0.85\linewidth]{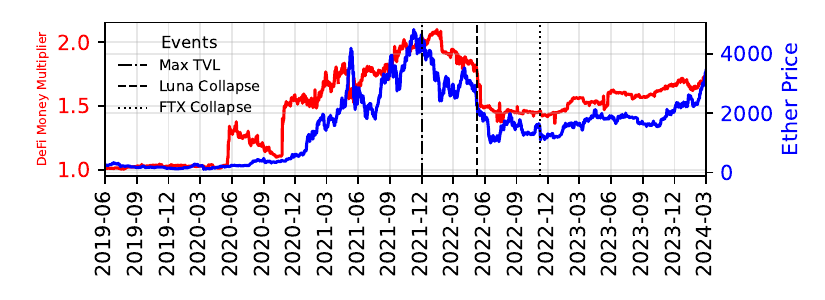}
    \vspace{-5mm}
    \caption{\ac{defi} money multiplier (red line) and \ac{eth} price (blue line).}
    \label{fig:money_supply_compare}
\end{figure}

\subsection{DeFi Money Multiplier}

\label{subsec:defi_money_multiplier}

The M2 to M0 ratio, known as the traditional \href{https://fredblog.stlouisfed.org/2023/07/the-monetary-multiplier-and-bank-reserves/}{money multiplier}, indicates the extent to which banks can utilize investor deposits. M0 denotes the money base, which includes cash and bank reserves. M2 denotes the supply of private money, which includes cash, checking deposits, and other short-term deposits. Drawing a parallel, we can divide \ac{tvl} by \ac{tvr} to compute the \ac{defi} money multiplier.
This ratio reflects the degree of double counting and wrapping effects within the \ac{defi} ecosystem, analogous to the money multiplier in \ac{trafi}. \autoref{fig:money_supply_compare} plots the \ac{defi} money multiplier.

\autoref{tab:corr} lists the Spearman's rank correlation coefficients between the \ac{defi} money multiplier ($M^{\textnormal{DeFi}}$), key macroeconomic indicators in the US, and representative crypto market indicators. Notably, there is a significant positive correlation between the \ac{defi} money multiplier and cryptocurrency market indicators, such as the S\&P Cryptocurrency Broad Digital Market Index (${\it S\&P}$) and Ethereum price (${\it ETH}$). This suggests that during bullish periods in the cryptocurrency market, investors tend to increase their investments in \ac{defi} and actively engage in leveraged positions. Conversely, the \ac{defi} money multiplier is significantly negatively correlated with the \ac{trafi} money multiplier ($M^{\textnormal{TradFi}}$). However, the \ac{defi} money multiplier does not exhibit a significant correlation with the Consumer Price Index (${\it CPI}$) or the CBOE Volatility Index (${\it VIX}$). As a robustness test, we also calculate Spearman's rank correlation coefficients between the natural logarithmic return of these indicators to make variables stationary, which is shown in Appendix~\autoref{sec:log_corr}.

\begin{table}[tb]
    \centering
    \tiny
    \renewcommand{\arraystretch}{0.5}
    \caption{Spearman's rank correlation coefficients between macroeconomic indicators, cryptocurrency market indicators, and \ac{defi} money multiplier computed from \ac{tvl} and \ac{tvr}.}
    \begin{tabularx}{\linewidth}{>{\raggedright\arraybackslash}p{1cm}*{6}{X}}
\toprule
 & \multicolumn{3}{c}{Macroeconomic / TradFi indicators} & \multicolumn{3}{c}{Cryptocurrency / DeFi indicators}\\
 \cmidrule(lr){2-4}\cmidrule(lr){5-7}
 & ${\it CPI_t}$ & ${\it VIX_t}$ & ${\it M_t^{\textnormal{TradFi}}}$ & ${\it ETH_t}$ & ${\it S\&P_t}$ & ${\it M_t^{\textnormal{DeFi}}}$ \\
${\it CPI_t}$ & {\cellcolor[HTML]{FF8080}} \color[HTML]{F1F1F1} 1.00$^{***}$ & {\cellcolor[HTML]{E8E8FF}} \color[HTML]{000000} -0.19 & {\cellcolor[HTML]{FFF6F6}} \color[HTML]{000000} 0.06 & {\cellcolor[HTML]{FFDBDB}} \color[HTML]{000000} 0.28$^{*}$ & {\cellcolor[HTML]{FFF6F6}} \color[HTML]{000000} 0.07 & {\cellcolor[HTML]{FFE0E0}} \color[HTML]{000000} 0.25 \\
${\it VIX_t}$ & {\cellcolor[HTML]{E8E8FF}} \color[HTML]{000000} -0.19 & {\cellcolor[HTML]{FF8080}} \color[HTML]{F1F1F1} 1.00$^{***}$ & {\cellcolor[HTML]{ECECFF}} \color[HTML]{000000} -0.15 & {\cellcolor[HTML]{E4E4FF}} \color[HTML]{000000} -0.21 & {\cellcolor[HTML]{E6E6FF}} \color[HTML]{000000} -0.19 & {\cellcolor[HTML]{E8E8FF}} \color[HTML]{000000} -0.18 \\
${\it M_t^{\textnormal{TradFi}}}$ & {\cellcolor[HTML]{FFF6F6}} \color[HTML]{000000} 0.06  & {\cellcolor[HTML]{ECECFF}} \color[HTML]{000000} -0.15 & {\cellcolor[HTML]{FF8080}} \color[HTML]{F1F1F1} 1.00$^{***}$ & {\cellcolor[HTML]{ABABFF}} \color[HTML]{000000} -0.66$^{***}$ & {\cellcolor[HTML]{A6A6FF}} \color[HTML]{000000} -0.70$^{***}$ & {\cellcolor[HTML]{9E9EFF}} \color[HTML]{F1F1F1} -0.76$^{***}$ \\
${\it ETH_t}$ & {\cellcolor[HTML]{FFDBDB}} \color[HTML]{000000} 0.28$^{*}$ & {\cellcolor[HTML]{E4E4FF}} \color[HTML]{000000} -0.21 & {\cellcolor[HTML]{ABABFF}} \color[HTML]{000000} -0.66$^{***}$ & {\cellcolor[HTML]{FF8080}} \color[HTML]{F1F1F1} 1.00$^{***}$ & {\cellcolor[HTML]{FF8686}} \color[HTML]{F1F1F1} 0.95$^{***}$ & {\cellcolor[HTML]{FF8B8B}} \color[HTML]{000000} 0.91$^{***}$ \\
\bottomrule
\end{tabularx}
    \raggedright ***, **, and * denote the 1\%, 5\%, and 10\% significance levels, respectively.
    \label{tab:corr}
    \vspace{-5mm}
\end{table} 
\section{Risk Analyses}
\label{sec:risk_analysis}

In this section, we present the outcomes of the comparative sensitivity tests. Using the data for the six leading \ac{defi} protocols in \autoref{fig:complexity}, these tests are conducted on three representative date snapshots, each with a different set of parameters.
We discuss the default value of certain parameters in Appendix~\autoref{sec:params}.
To provide an overview of the simulation environment, we visualize the wrapping network of these protocols in \autoref{fig:protocol_network}. From this visualization, we observe that both \ac{tvl} and \ac{tvl}-to-\ac{tvr} ratio $\frac{{\it TVL}}{{\it TVR}}$ of protocols excluding Lido decreases from the point when \ac{tvl} reaches the maximum value to the subsequent collapse of LUNA and FTX. This trend suggests a reduction in both the overall size of the system and the extent of double-counting within it, which are aligned with the broader dynamics of the overall \ac{defi} system,  as depicted in \autoref{fig:tvl_Total} and \autoref{fig:tvl_decompose_sankey}. 

\begin{figure*}[tb]
\centering

\begin{subfigure}{0.31\linewidth}
    \centering
    \tikz\node[inner sep=0pt,
           label=west:\rotatebox{90}{\tiny\textbf{Max \ac{tvl}}}]{\includegraphics[trim=0 36pt 0 0, clip, width=\linewidth]{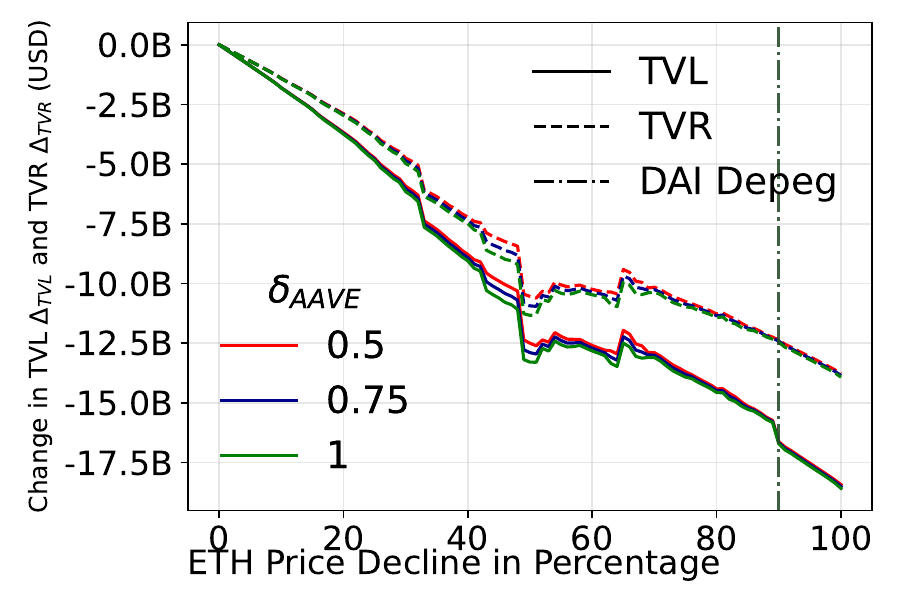}};
    \label{subfig:sensitivity_max_tvl_aave_c}
\end{subfigure}%
\hfill
\begin{subfigure}{0.31\linewidth}
    \centering
    \includegraphics[trim=0 36pt 0 0, clip, width=\linewidth]{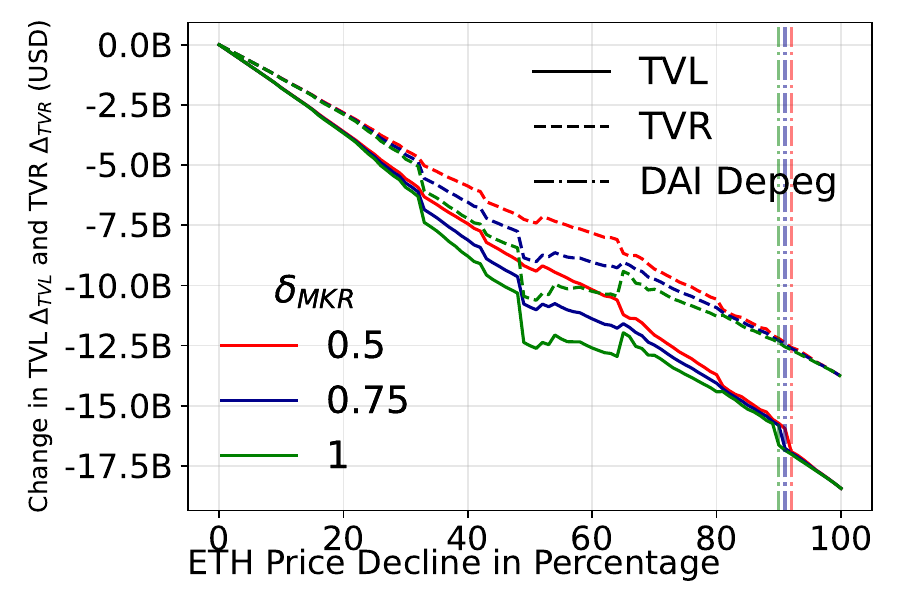}
    \label{subfig:sensitivity_max_tvl_mkr_c}
\end{subfigure}%
\hfill
\begin{subfigure}{0.31\linewidth}
    \centering
    \includegraphics[trim=0 36pt 0 0, clip, width=\linewidth]{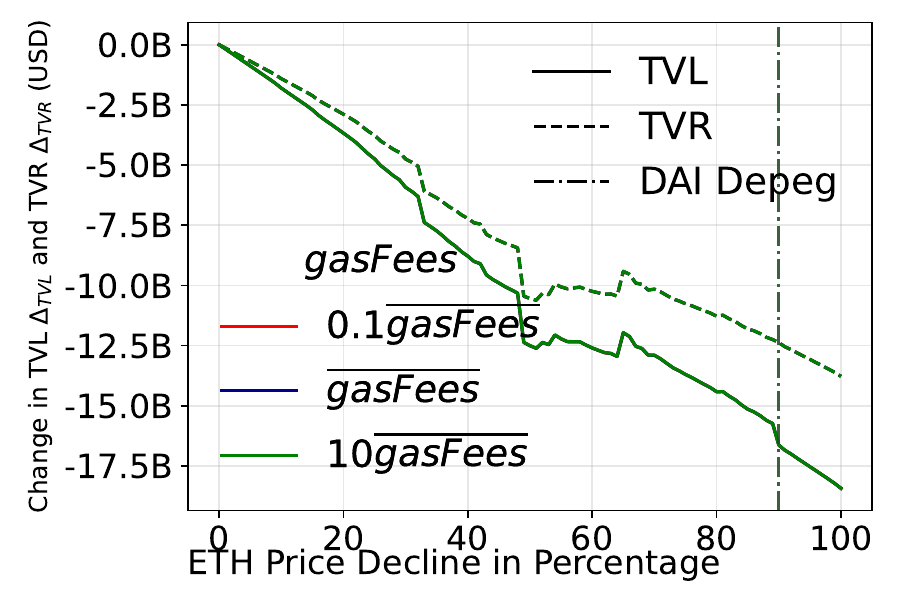}
    \label{subfig:sensitivity_max_tvl_gas}
\end{subfigure}
\vfill
\begin{subfigure}{0.31\linewidth}
    \centering
    \tikz\node[inner sep=0pt,
           label=west:\rotatebox{90}{\tiny\textbf{Luna Collapse}}]
    {\includegraphics[trim=0 36pt 0 0, clip, width=\linewidth]{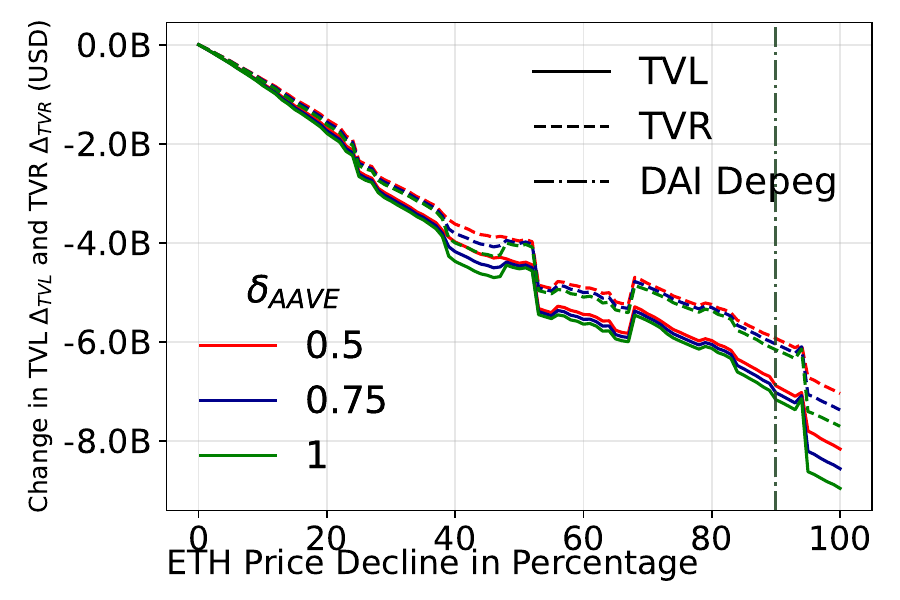}};
    \label{subfig:sensitivity_luna_collapse_aave_c}
\end{subfigure}%
\hfill
\begin{subfigure}{0.31\linewidth}
    \centering
    \includegraphics[trim=0 36pt 0 0, clip, width=\linewidth]{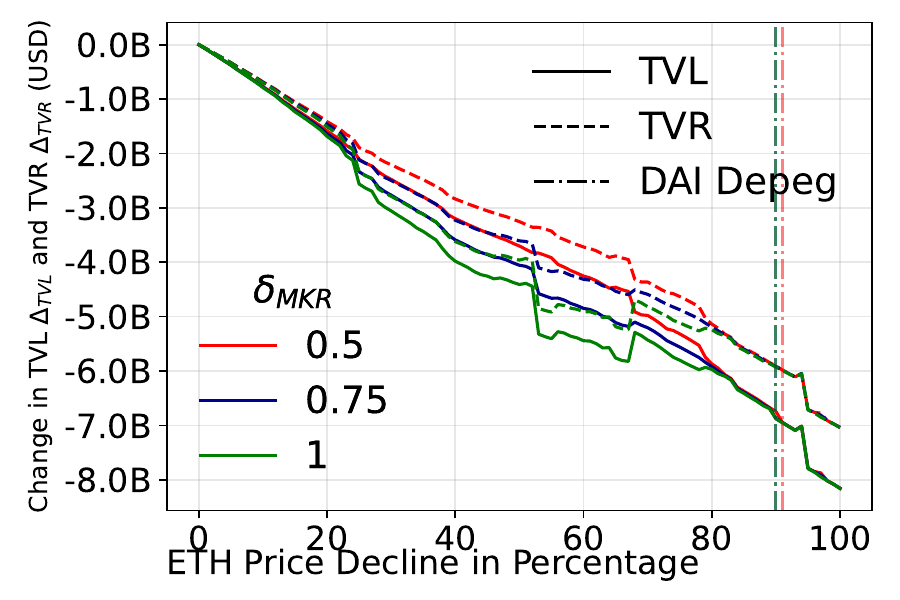}
    \label{subfig:sensitivity_luna_collapse_mkr_c}
\end{subfigure}%
\hfill
\begin{subfigure}{0.31\linewidth}
    \centering
    \includegraphics[trim=0 36pt 0 0, clip, width=\linewidth]{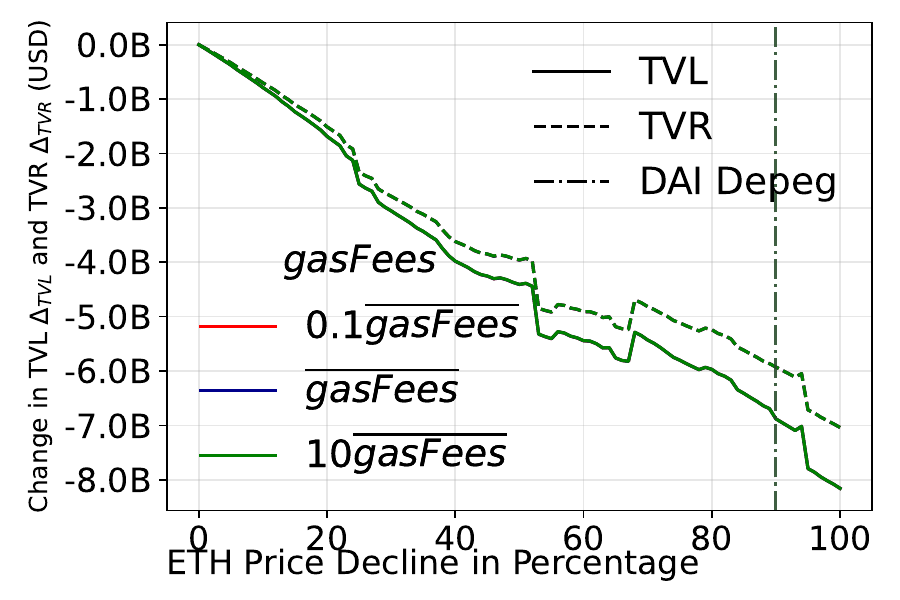}
    \label{subfig:sensitivity_luna_collapse_gas}
\end{subfigure}
\vfill
\begin{subfigure}{0.31\linewidth}
    \centering
    \tikz\node[inner sep=0pt,
           label=west:\rotatebox{90}{\tiny\textbf{FTX Collapse}},
           label=below:\tiny\textbf{Close Factor, Aave V2}]
    {\includegraphics[width=\linewidth]{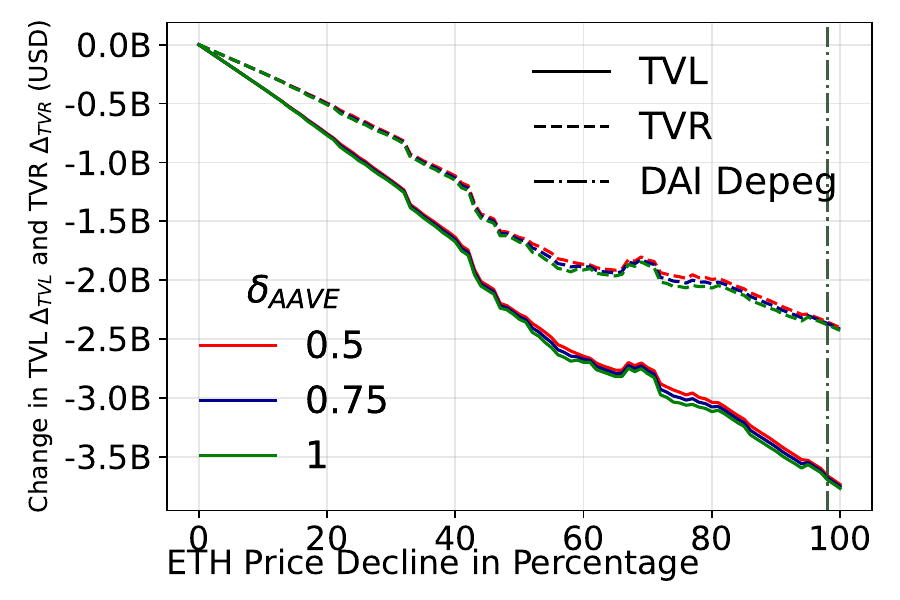}};
    \label{subfig:sensitivity_ftx_collapse_aave_c}
\end{subfigure}%
\hfill
\begin{subfigure}{0.31\linewidth}
    \centering
    \tikz\node[inner sep=0pt,
           label=below:\tiny\textbf{Close Factor, MakerDAO}]
    {\includegraphics[width=\linewidth]{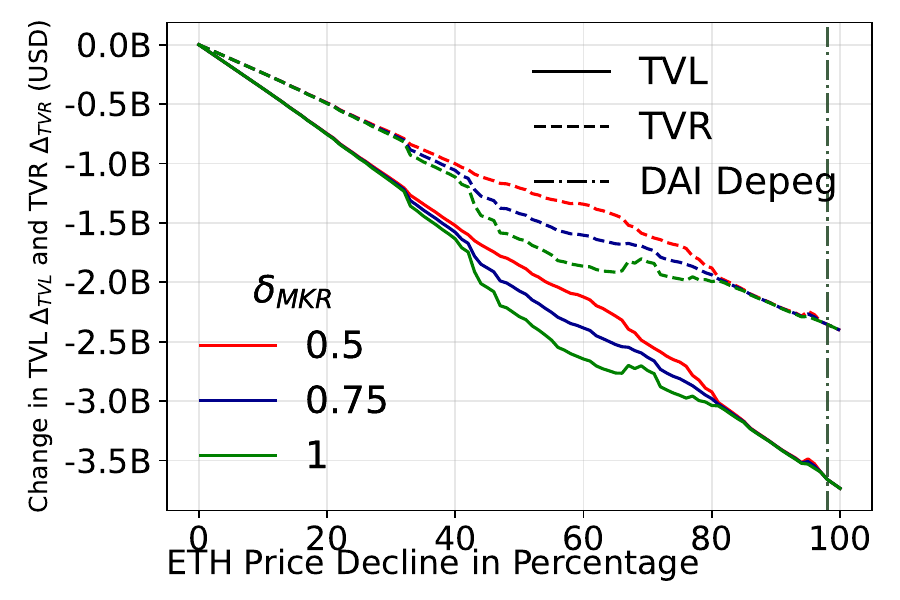}};
    \label{subfig:sensitivity_ftx_collapse_mkr_c}
\end{subfigure}%
\hfill
\begin{subfigure}{0.31\linewidth}
    \centering
    \tikz\node[inner sep=0pt,
           label=below:\tiny\textbf{Gas Fees}]
    {\includegraphics[width=\linewidth]{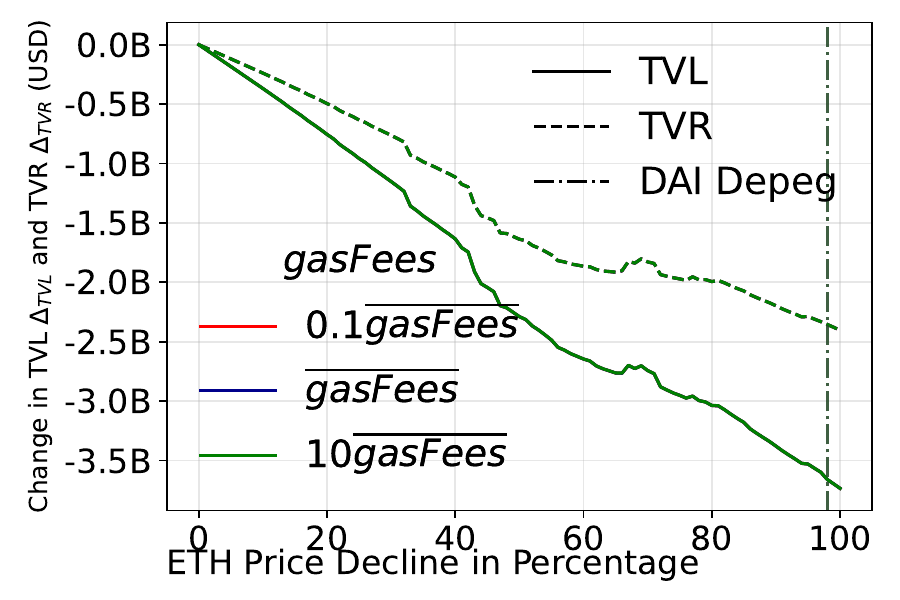}};
    \label{subfig:sensitivity_ftx_collapse_gas}
\end{subfigure}\\
\vspace{-5mm}
\caption{Change in \ac{tvl} $\Delta_{\it TVL}$ and \ac{tvr} $\Delta_{\it TVR}$ as a function of ETH price decline in percentage $d$ on three representative point with different parameter values. Subfigures within a row represent sensitivity tests conducted under the same snapshot, whereas subfigures within a column display sensitivity tests implemented under three different sets of values for a given parameter. Vertical dashed lines indicate the timing of the DAI depeg under different parameter settings.}
\label{fig:sensitivity_test}
\end{figure*}

\autoref{fig:sensitivity_test} shows how $\Delta {\it TVL}$ and $\Delta {\it TVR}$ vary with $d_{\textnormal{ETH}}$. The $\Delta {\it TVL}$ and $\Delta {\it TVR}$ curves with the default parameter setting in Appendix~\autoref{sec:params} are compared with those with hypothetical parameters. Irrespective of parameter values, the $\Delta {\it TVL}$ curve is more sensitive to $d_{\textnormal{ETH}}$ than the $\Delta {\it TVR}$ curve due to the financial contagion effect of the derivative tokens, which aligns with the reasoning in \autoref{sec:enhanced_measurement_framework}.

We then discuss how other parameters in \autoref{eq:endo_quantity} affects $\Delta {\it TVL}$ and $\Delta {\it TVR}$:
\vspace{-3.5mm}
\begin{enumerate}
    \item Close factor ($\delta$): The close factor $\delta$ in a lending protocol represents the portion of a loan that a liquidator is allowed to repay when a borrower's health factor falls below one. For instance, consider a liquidable loan position where the loan amount is \$100 and the close factor is 0.8; the maximum amount that can be liquidated is then \$80 worth of tokens. As shown in \autoref{fig:sensitivity_test}, higher $\delta$ leads to a greater drop in $\Delta {\it TVL}$ and $\Delta {\it TVR}$, ceteris paribus. This effect is observed for both $\delta_{\textnormal{AAVE}}$ and $\delta_{\textnormal{MKR}}$. The drop in $\Delta {\it TVL}$ is more sensitive to $\delta_{\textnormal{MKR}}$ than $\delta_{\textnormal{AAVE}}$ since the MakerDAO has greater exposure to a decline of \ac{eth} price compared to Aave V2.

    \item Gas fees ($\overline{gasFees}$). We first calculate the average gas fee across daily transactions using $\overline{gasFees} = \overline{gasLimit \times gasPrice}$, and then adjust it by scaling with factors of 0.1, 1, and 10. As shown in \autoref{fig:sensitivity_test}, the variations in $\overline{gasFees}$ have a minimal impact on $\Delta {\it TVL}$ and $\Delta {\it TVR}$. This indicates that transaction fees are generally negligible compared to the change of collateral value for most positions.
    
\end{enumerate}
\section{Related Work}
\label{sec:related_work}

Several studies explore the role of \ac{tvl} in \ac{defi} valuation and risk monitoring. Metelski et al.~\cite{Kumar2022DecentralizedValuations} and Xu et al.~\cite{Xu2022e} investigate the causal relationship between key \ac{defi} performance metrics, such as \ac{tvl}, protocol revenues, and the \ac{defi} protocol valuations. Stepanova et al.~\cite{Stepanova2021ReviewLocked} conduct preliminary descriptive and comparison work on \ac{tvl} of 12 most popular \ac{defi} protocols. Maouchi et al.\cite{Maouchi2022UnderstandingNFTs} show that \ac{tvl} can work as a valuable tool for monitoring market dynamics and assessing the risk of bubbles in the digital financial landscape. {\c{S}}oiman et al.~\cite{Soiman2023WhatReturns} use
\ac{tvl} divided by market capitalization for \ac{defi} valuation and examine whether this metric drives the \ac{defi} returns.

Some studies examine the \ac{defi} composability and \ac{tvl} double counting problem in a limited scope. Kitzler et al.~\cite{Kitzler2023DisentanglingCompositions} measure the composition of \ac{defi} protocols. Saengchote~\cite{Saengchote2021WhereMeans.} examines the flow of DAI, a \ac{defi} stablecoin, between protocols using high-frequency transaction-level data. The study also explains how \ac{tvl} accounts for repeat value through the wrapping of DAI. Chiu et al.~\cite{Chiu2023UnderstandingModel} use a standard theoretical production-network model to assess the value added and service outputs across various \ac{defi} sectors on Ethereum.

\section{Conclusion}
\label{sec:concl}

This paper presents a novel yet effective measurement framework, \ac{tvr}, to thoroughly address the double counting problem in \ac{defi}. We find a substantial amount of double counting within the DeFi system. Our sensitivity tests show that \ac{tvl} is highly sensitive during market downturns. We also document that the DeFi money multiplier is positively correlated with crypto market indicators and negatively correlated with macroeconomic indicators. Overall, our findings suggest that \ac{tvr} is more reliable and stable than \ac{tvl}.

\section*{Acknowledgments}

This material is based upon work partially supported by Ripple under the University Blockchain Research Initiative (UBRI)~\cite{Feng2022UniversityResearch}. Any opinions, findings, and conclusions or recommendations expressed in this material are those of the authors and do not necessarily reflect the views of Ripple.

We thank Antoine Mouran for his insights on the double counting problem and related literature. We also appreciate the valuable comments provided by Carol Alexander, Ka Wai Ng, Kamil Tylinski, Wenzhi Ding, Francesca Medda, Chao Liu, Hui Gong, Walter Hernandez Cruz, Yitian Wang, and Honglin Fu.

\bibliographystyle{splncs04}
\bibliography{references}

\begin{thebibliography}{10}
\providecommand{\url}[1]{\texttt{#1}}
\providecommand{\urlprefix}{URL }
\providecommand{\doi}[1]{https://doi.org/#1}

\bibitem{Andolfatto2017RehypothecationLiquidity}
Andolfatto, D., Martin, F.M., Zhang, S.: {Rehypothecation and liquidity}. European Economic Review  \textbf{100},  488--505 (11 2017). \doi{10.1016/J.EUROECOREV.2017.09.010}

\bibitem{Arora2024SecPLF:Attacks}
Arora, S., Li, Y., Feng, Y., Xu, J.: {SecPLF: Secure Protocols for Loanable Funds against Oracle Manipulation Attacks}. In: The 19th ACM Asia Conference on Computer and Communications Security. vol.~12, pp. 1394--1405. ACM, New York, NY, USA (7 2024). \doi{10.1145/3634737.3637681}

\bibitem{Baird2018Hedera:Council}
Baird, L., Harmon, M., Madsen, P.: {Hedera: A Public Hashgraph Network {\&} Governing Council}  (2018)

\bibitem{Chiu2023UnderstandingModel}
Chiu, J., Koeppl, T.V., Yu, H., Zhang, S.: {Understanding Defi Through the Lens of a Production-Network Model}. SSRN Electronic Journal  (6 2023). \doi{10.2139/SSRN.4487615}

\bibitem{Clevenger1943PresentingCredit}
Clevenger, E.: {Presenting the Theory of Debit and Credit}. The Accounting Review  \textbf{18}(1),  40--44 (1 1943), \url{https://www.jstor.org/stable/240361}

\bibitem{Cousaert2021}
Cousaert, S., Xu, J., Matsui, T.: {SoK: Yield Aggregators in DeFi}. In: ICBC. pp. 1--14. IEEE (5 2022). \doi{10.1109/ICBC54727.2022.9805523}

\bibitem{Feng2022UniversityResearch}
Feng, Y., Xu, J., Weymouth, L.: {University Blockchain Research Initiative (UBRI): Boosting blockchain education and research}. IEEE Potentials  \textbf{41}(6),  19--25 (11 2022). \doi{10.1109/MPOT.2022.3198929}

\bibitem{Gogol2024SoK:Risks}
Gogol, K., Killer, C., Schlosser, M., Bocek, T., Stiller, B., Tessone, C.: {SoK: Decentralized Finance (DeFi) -- Fundamentals, Taxonomy and Risks} pp. 1--20 (4 2024). \doi{10.48550/arXiv.2404.11281}

\bibitem{Hedera2024SendContracts}
{Hedera}: {Send and Receive HBAR Using Solidity Smart Contracts} (2024), \url{https://docs.hedera.com/hedera/tutorials/smart-contracts/send-and-receive-hbar-using-solidity-smart-contracts}

\bibitem{Hernandez2025EvolutionLiterature}
Hernandez, W., Tylinski, K., Moore, A., Roche, N., Vadgama, N., Treiblmaier, H., Shangguan, J., Tasca, P., Xu, J.: {Evolution of ESG-focused DLT Research: An NLP Analysis of the Literature} (8 2025), \url{https://doi.org/10.48550/arXiv.2308.12420}

\bibitem{Ibanez2020c}
Iba{\~{n}}ez, J.I., Bayer, C.N., Tasca, P., Xu, J.: {REA, Triple-Entry Accounting and Blockchain: Converging Paths to Shared Ledger Systems}. Journal of Risk and Financial Management  \textbf{16}(9), ~382 (8 2023). \doi{10.3390/jrfm16090382}

\bibitem{Ibanez2025Triple-EntryKin}
Iba{\~{n}}ez, J.I., Bayer, C.N., Tasca, P., Xu, J.: {Triple-Entry Accounting, Blockchain, and Next of Kin}. In: Aggarwal, R., Tasca, P. (eds.) Digital Assets, chap.~9, pp. 198--226. Cambridge University Press (2 2025). \doi{10.1017/9781009362290.012}

\bibitem{Kitzler2023DisentanglingCompositions}
Kitzler, S., Victor, F., Saggese, P., Haslhofer, B.: {Disentangling Decentralized Finance (DeFi) Compositions}. ACM Transactions on the Web  \textbf{17}(2), ~10 (3 2023). \doi{10.1145/3532857}

\bibitem{Kumar2022DecentralizedValuations}
Kumar, A., Le, T., Elnahass, M., Metelski, D., Sobieraj, J.: {Decentralized Finance (DeFi) Projects: A Study of Key Performance Indicators in Terms of DeFi Protocols’ Valuations}. International Journal of Financial Studies  \textbf{10}(4), ~108 (11 2022). \doi{10.3390/IJFS10040108}

\bibitem{MakerDAO2022PegStability}
{MakerDAO}: {Peg Stability} (2022), \url{https://manual.makerdao.com/module-index/module-psm}

\bibitem{Maouchi2022UnderstandingNFTs}
Maouchi, Y., Charfeddine, L., El~Montasser, G.: {Understanding digital bubbles amidst the COVID-19 pandemic: Evidence from DeFi and NFTs}. Finance Research Letters  \textbf{47},  102584 (6 2022). \doi{10.1016/J.FRL.2021.102584}

\bibitem{Panasyuk2021DeFiDECENTRALIZATION}
Panasyuk, V., Brechko, O., Buchynska, T.: {DeFi Experiment as a Form of Cryptocurrencies Market Transformation and In-depth Decentralization}  (2021), \url{http://dspace.wunu.edu.ua//handle/316497/46201}

\bibitem{Perez2021Liquidations:Knife-Edge}
Perez, D., Werner, S.M., Xu, J., Livshits, B.: {Liquidations: DeFi on a Knife-Edge}. In: FC. vol. 12675, pp. 457--476. Springer (2021), \url{https://doi.org/10.1007/978-3-662-64331-0_24}

\bibitem{Perez2020}
Perez, D., Xu, J., Livshits, B.: {Revisiting Transactional Statistics of High-scalability Blockchains}. In: IMC. vol.~16, pp. 535--550. ACM, New York (10 2020). \doi{10.1145/3419394.3423628}

\bibitem{Qin2021AnLiquidations}
Qin, K., Zhou, L., Gamito, P., Jovanovic, P., Gervais, A.: {An empirical study of DeFi liquidations}. In: ACM Internet Measurement Conference. pp. 336--350. ACM, New York, NY, USA (11 2021). \doi{10.1145/3487552.3487811}

\bibitem{Richardson2020e}
Richardson, A., Xu, J.: {Carbon Trading with Blockchain}. In: Pardalos, P., Kotsireas, I., Guo, Y., Knottenbelt, W. (eds.) MARBLE, chap.~7, pp. 105--124 (5 2020), \url{https://doi.org/10.1007/978-3-030-53356-4_7}

\bibitem{Saengchote2021WhereMeans.}
Saengchote, K.: {Where do DeFi Stablecoins Go? A Closer Look at What DeFi Composability Really Means.} SSRN Electronic Journal  (7 2021). \doi{10.2139/SSRN.3893487}

\bibitem{Simmonds1986ConsolidatedSubsidiary}
Simmonds, A.: {Consolidated Balance Sheet—Direct Subsidiary}. In: Mastering Financial Accounting, pp. 295--318. Palgrave, London (1986), \url{https://doi.org/10.1007/978-1-349-18430-9_17}

\bibitem{Soiman2023WhatReturns}
{\c{S}}oiman, F., Dumas, J.G., Jimenez-Garces, S.: {What drives DeFi market returns?} Journal of International Financial Markets, Institutions and Money  \textbf{85},  101786 (6 2023). \doi{10.1016/J.INTFIN.2023.101786}

\bibitem{Spearman1904TheThings}
Spearman, C.: {The Proof and Measurement of Association between Two Things}. The American Journal of Psychology  \textbf{15}(1), ~72 (1 1904). \doi{10.2307/1412159}

\bibitem{Stepanova2021ReviewLocked}
Stepanova, V., Eriņ{\v{s}}, I.: {Review of Decentralized Finance Applications and Their Total Value Locked}. TEM Journal  \textbf{10}(1),  327--333 (2021). \doi{10.18421/TEM101-41}

\bibitem{Tovanich2023ContagionCompound}
Tovanich, N., Kassoul, M., Weidenholzer, S., Prat, J.: {Contagion in Decentralized Lending Protocols: A Case Study of Compound}. In: Workshop on Decentralized Finance and Security. pp. 55--63. ACM, New York (11 2023). \doi{10.1145/3605768.3623544}

\bibitem{Ultron2023UltronWhitepaper}
{Ultron}: {Ultron Whitepaper}  (2023)

\bibitem{Xu2022g}
Xu, J., Feng, Y.: {Reap the Harvest on Blockchain: A Survey of Yield Farming Protocols}. IEEE Transactions on Network and Service Management  \textbf{20}(1),  858--869 (3 2023). \doi{10.1109/TNSM.2022.3222815}

\bibitem{Xu2025Auto.gov:DeFi}
Xu, J., Feng, Y., Perez, D., Livshits, B.: {Auto.gov: Learning-based Governance for Decentralized Finance (DeFi)}. IEEE Transactions on Services Computing pp. 1278--1292 (2 2025). \doi{10.1109/TSC.2025.3553700}

\bibitem{Xu2023SoK:Protocols}
Xu, J., Paruch, K., Cousaert, S., Feng, Y.: {SoK: Decentralized Exchanges (DEX) with Automated Market Maker (AMM) Protocols}. ACM Computing Surveys  \textbf{55}(11) (2 2023). \doi{10.1145/3570639}

\bibitem{Xu2022d}
Xu, J., Vadgama, N.: {From Banks to DeFi: the Evolution of the Lending Market}. In: Vadgama, N., Xu, J., Tasca, P. (eds.) Enabling the Internet of Value, chap.~6, pp. 53--66. Springer, Cham (1 2022), \url{https://doi.org/10.1007/978-3-030-78184-2_6}

\bibitem{Xu2022b}
Xu, T.A., Xu, J.: {A Short Survey on Business Models of Decentralized Finance (DeFi) Protocols}. In: FC. pp. 197--206. Springer (2023), \url{https://doi.org/10.1007/978-3-031-32415-4_13}

\bibitem{Xu2022e}
Xu, T.A., Xu, J., Lommers, K.: {DeFi versus TradFi Valuation Using Multiples and Discounted Cash Flows}. In: Aggarwal, R., Tasca, P. (eds.) Digital Assets, pp. 44--68. Cambridge University Press (2 2025). \doi{10.1017/9781009362290.004}

\end{thebibliography}

\appendix
\section*{APPENDIX}





\section{Leveraging}

\label{subsec:leveraging}

\begin{figure}[b]
\centering
\begin{subfigure}{.5\linewidth}
  \centering
  \includegraphics[width=0.9\linewidth]{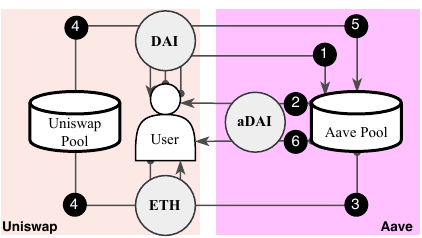}
  \caption{Leveraging in \ac{defi}.}
  \label{subfig:leverage_defi}
\end{subfigure}%
\hfill
\begin{subfigure}{.5\linewidth}
  \centering
  \includegraphics[width=0.9\linewidth]{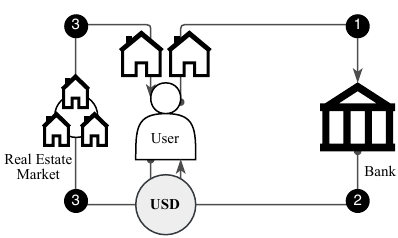}
  \caption{Leveraging in TradFi.}
  \label{subfig:leverage_tradfi}
\end{subfigure}
\caption{Leveraging and its corresponding \ac{trafi} analogies. Both \ac{defi} and \ac{trafi} involve leveraging processes, depicted in \autoref{subfig:leverage_defi} and \autoref{subfig:leverage_tradfi}, respectively. The black circle (\CIRCLE) with a white number indicates the step.}
\label{fig:double-counting-2}
\end{figure}

Leveraging is the other mechanism that can lead to \ac{tvl} double counting. It means investing with borrowed assets. The leveraging operation in \ac{defi} is similar to that in \ac{trafi}. In the \ac{trafi} leveraging scenario illustrated in \autoref{subfig:leverage_tradfi}, the investor can use her house as collateral (step \encircle{1}) to borrow cash (step \encircle{2}) and then use cash to buy another house (step \encircle{3}). In the \ac{defi} leveraging scenario illustrated in \autoref{subfig:leverage_defi}, we assume that liquidity providers have already supplied \$900 ETH to Aave and \$1.8k ETH along with \$1.8k DAI to Uniswap to facilitate swaps and borrowing (step \encircle{0}). Initially, Aave has \$900 ETH in assets and \$900 worth of aETH in liabilities, while Uniswap holds \$1.8k ETH and \$1.8k DAI in assets and \$1.8k worth of ETH-DAI LP tokens in liabilities. An investor provides \$2k DAI (step \encircle{1}) as collateral to borrow \$900 \ac{eth} (step \encircle{3}) in Aave, swaps the borrowed \$900 \ac{eth} to \$900 DAI (step \encircle{4}) in Uniswap, and then deposits \$900 DAI (step \encircle{5}) from Uniswap in the Aave to issue the receipt token aDAI (step 6).

\autoref{tab:consolidated-balance-sheet-2} shows the balance sheets of Aave and Uniswap, and the consolidated balance sheet of the \ac{defi} system consisting of Aave and Uniswap. In $\mathcal{S}^{(1)}_{\textnormal{L}}$, the value of the DeFi system is \$5.6k. However, if the user borrows \$900 \ac{eth}, swaps the \ac{eth} into DAI, and deposits the DAI into Aave, the \ac{tvl} will be \$6.5k under the traditional \ac{tvl} measurement. This \ac{tvl} is also double-counted because it includes DAI. In the consolidated balance sheet, the \ac{tvl} is adjusted to \$5.6k after eliminating the value associated with DAI.

\begin{table}[tb]
\centering
\tiny
\caption{Protocol-perspective balance sheets of the leveraging scenario. We highlight the components included in the \ac{tvl} calculation in red.}

\label{tab:consolidated-balance-sheet-2}
\begin{subtable}{0.32\linewidth}
\caption{The balance sheet of Aave.}
\label{subtab:balance-sheet-curve}
\begin{tabular}{@{}p{0.63\linewidth}>{\raggedleft\arraybackslash}p{0.14\linewidth}>{\raggedleft\arraybackslash}p{0.17\linewidth}@{}}
\toprule
\textbf{Immediately After} & \encircle{3} & \encircle{5} \\
\midrule
\textit{Assets} & \$ & \$ \\
Value Locked - DAI & \textcolor{red}{2,000} & \textcolor{red}{2,900} \\
Receivables - dETH & 900 & 900 \\
\midrule
\textbf{Total Assets} & 2,900 & 3,800 \\
\bottomrule\bottomrule
\textit{Liabilities} & \$ & \$ \\
Payables - aDAI & 2,000 & 2,900 \\
Payables - aETH & 900 & 900\\
\midrule
\textbf{Total Liabilities}  & 2,900 & 3,800 \\
\bottomrule\bottomrule
\end{tabular}
\end{subtable}%
\hfill
\begin{subtable}{0.32\linewidth}
\caption{The balance sheet of Uniswap.}
\label{subtab:balance-sheet-yearn}
\begin{tabular}{@{}p{0.65\linewidth}>{\raggedleft\arraybackslash}p{0.14\linewidth}>{\raggedleft\arraybackslash}p{0.17\linewidth}@{}}
\toprule
\textbf{Immediately After} & \encircle{3} & \encircle{5} \\
\midrule
\textit{Assets} & \$ & \$ \\
Value Locked - ETH & \textcolor{red}{1,800} & \textcolor{red}{2,700} \\
Value Locked - DAI & \textcolor{red}{1,800} & \textcolor{red}{900} \\
\midrule
\textbf{Total Assets} & 3,600 & 3,600 \\
\bottomrule\bottomrule
\textit{Liabilities} & \$ & \$ \\
Payables - ETH-DAI-LP & 3,600 & 3,600 \\
&&\\
\midrule
\textbf{Total Liabilities} & 3,600 & 3,600 \\
\bottomrule\bottomrule
\end{tabular}
\end{subtable}%
\hfill
\begin{subtable}{0.33\linewidth}
\caption{Consolidated balance sheet.}
\label{subtab:consolidated-balance-sheet-2}
\begin{tabular}{@{}p{0.63\linewidth}>{\raggedleft\arraybackslash}p{0.14\linewidth}>{\raggedleft\arraybackslash}p{0.17\linewidth}@{}}
\toprule
\textbf{Immediately After} & \encircle{3} & \encircle{5} \\
\midrule
\textit{Assets} & \$ & \$ \\
Value Locked & \textcolor{red}{5,600} & \textcolor{red}{5,600} \\
&&\\
\midrule
\textbf{Total Assets} & 5,600 & 5,600 \\

\bottomrule\bottomrule
\textit{Liabilities} & \$ & \$ \\
Payables & 5,600 & 5,600 \\
&&\\
\midrule
\textbf{Total Liabilities} & 5,600 & 5,600 \\
\bottomrule\bottomrule
\end{tabular}
\end{subtable}
\end{table}

\section{DeFi Bookkeeping}
\label{sec:defi_bookkeeping}
We model value transfers of common transactions in \ac{defi} protocols via the double-entry bookkeeping of the traditional accounting framework in \autoref{tab:journalization-makerdao}. The double-entry booking keeping has two equal journal entries known as debit (Dr.) and credit (Cr.), which represent a transfer of value to and from that account, respectively~\cite{Clevenger1943PresentingCredit}. The credit side should be indented. 

\begin{table}[!ht]
\centering
\caption{Journal entries of common transactions in \ac{defi} protocols.}
\resizebox{\textwidth}{!}{
\tiny
\begin{tabular}{m{1.2cm}m{1.5cm}m{3cm}m{3cm}m{3.3cm}lm{1.2cm}}
\toprule

\textbf{Protocol} & \textbf{Transaction} & \textbf{Definition} & \textbf{Journal Entry} & \textbf{Amount} & \textbf{Type} & \textbf{Example Transaction}\\

\midrule

Lido &
Staking. &
User stakes $q_T$ tokens T with dollar price $p_{T}$ to issue $q_R$ receipt token R with dollar price $p_{R}$.&
\makecell[l]{Dr. Value Locked -- T \\ \hspace{3mm} Cr. Payables -- R} & 
\makecell[l]{$q_T p_{T}$ \\ \hspace{3mm} $q_R p_{R}$ } & 
\makecell[l]{Assets \\ Liabilities} &
\href{https://etherscan.io/tx/0x85d3538f6b34c184f57937e190132af38ec0b4442969d41013dd17ac3e05c697}{\makecell[l]{0x85d...c697}}\\

\cline{2-7}

&
Burning. &
User burns $q_R$ receipt token R with dollar price $p_{R}$ to redeem $q_T$ tokens T with dollar price $p_{T}$.&
\makecell[l]{Dr. Payables -- R \\ \hspace{3mm} Cr. Value Locked -- T} & 
\makecell[l]{$q_R  p_{R}$ \\ \hspace{3mm} $q_T  p_{T}$} & 
\makecell[l]{Liabilities \\ Assets} &
\href{https://etherscan.io/tx/0x2147c494d50049f3dce533aafa54cc9c0d5a7762057da2984dcf02125b921c6f}{\makecell[l]{0x214...1c6f}}\\

\midrule

MakerDAO &
Borrowing. & 
User supplies $q_T$ tokens T with dollar price $p_{T}$ as collateral to mint $q_S$ stablecoins S with dollar price $p_{S}$. &
\makecell[l]{Dr. Receivables -- S \\ Dr. Value Locked -- T \\  \hspace{3mm} Cr. New Money -- S \\ \hspace{3mm} Cr. Payables -- T} &
\makecell[l]{$q_S  p_{S}$ \\ $q_T  p_{T}$ \\ \hspace{3mm} $q_S  p_{S}$ \\ \hspace{3mm} $q_T  p_{T}$} &
\makecell[l]{Assets \\ Assets \\ Liabilities \\ Liabilities} &
\href{https://etherscan.io/tx/0x9594d2cb31496a6d75223d63aae65f4c77eb5028f5376132493f1a2e98880adb}{\makecell[l]{0x959...0adb}}\\
\bottomrule

\end{tabular}
}
\label{tab:journalization-makerdao}
\end{table}

\section{Overcollateralization and Liquidation}
\label{subsec:liquidation}

\Ac{plf} are \ac{defi} protocols that allow users to supply and borrow cryptocurrency~\cite{Qin2021AnLiquidations}. Analogous to a pawnshop, a \ac{plf} adheres to the overcollateralization mechanism, which means the borrower has to pledge collateral of higher total value than the loan~\cite{Panasyuk2021DeFiDECENTRALIZATION}. The ratio of the maximum amount of tokens that can be borrowed to the amount of collateral is referred to as the \ac{ltv}. \ac{plf} includes \ac{cdp} such as MakerDAO (orange block in \autoref{fig:complexity}) and lending protocols such as Aave (purple block). In contrast, Lido, Uniswap V2, Curve, and Convex are not considered a \ac{plf}. In a \ac{plf}, a user's account can have multiple collaterals and debts, which are called its position. We can express the quantities of collaterals of a position as a vector: $\mathbf{c} = \left[q_{i}\right]_{k \times 1}$, where $k$ are the number of collateral types. To convert token quantity into token value in USD, let $p$ denotes the collateral price. By integrating all collateral prices into a vector $\mathbf{p}_c = \left[p_{i}\right]_{k \times 1}$, we can express the value of an account's collateral as $\mathbf{c}^T \mathbf{p}_c$. Similarly, we can express the quantities, prices, and total value of debt of a position as $\mathbf{d}$, $\mathbf{p}_d$, and $\mathbf{d}^T \mathbf{p}_d$. Unlike a lending protocol, a \ac{cdp} uses staked collateral to issue stablecoins. Stablecoins are a type of crypto asset with a value pegged to a reference point, such as a fiat currency or another cryptocurrency.

During a market downturn, the position can be liquidated, resulting in a portion of the collateral being purchased by the liquidator at a discount when the liquidation is profitable. Whether an account is liquidated depends on its position's health ratio and liquidation profit. Each type of collateral has its own liquidation threshold $\alpha\in[0,1)$, which represents its borrowing capacity. Let $\bm{\alpha} = \left[\alpha_i\right]_{k \times 1}$ represents the vector of liquidation thresholds for all types of collaterals in a \ac{plf}. Health factor $h$ describes the financial health of an account, which can be expressed as \begin{equation}
    \label{eq:health_factor}
    h = \frac{\mathbf{c}^T \left(\bm{\alpha} \odot \mathbf{p}_c\right)}{\mathbf{d}^T \mathbf{p}_d}, %
\end{equation} %
where $\odot$ denotes the element-wise product. The position will be liquidated if $h\le1$ and the liquidation profit $\Pi$ is positive.

\section{Derivative Token Pegging Mechanism}
\label{derivative_token_pegging_mechanism}
\begin{table}[tb]
    \centering
    \small
    \caption{Derivative token pegging mechanism. Each subtable illustrates the strategy an arbitrager will adopt in the corresponding scenario. Different strategies for corresponding scenarios underpin the pricing formula \autoref{eq:endo_price}.}
    \begin{tabularx}{\linewidth}{l}
\toprule\toprule

\textbf{Arbitrager's Strategy 1} \quad $\epsilon_{d} < 0$, \hspace{0.3cm} $p_{d}' = p_{d} + \epsilon_{d}$\\
\midrule

1: \textbf{Buy} $\lambda$ derivative tokens $d$ at price $p_{d}'$ from the market using $\lambda \cdot p_{d}'$ (scalar $\lambda$\\
depends on the budget).\\

2: \textbf{Burn} derivative tokens $d$ to \textbf{redeem} all underlying tokens in the liquidity pool.\\

3: \textbf{Sell} all underlying tokens $d$ to get $\lambda \cdot p_{d}$ dollar and \textbf{increase} derivative token\\
price $p_{d}'$.\\

4: \textbf{Earn} profit $\lambda (p_{d} - p_{d}')$ and \textbf{repeat} steps 1 to 5 until $p_{d}' = p_{d}$, i.e. $\epsilon_{d} = 0$.\\
\bottomrule
\end{tabularx}
    \begin{tabularx}{\linewidth}{l}
\toprule

\textbf{Arbitrager's Strategy 2} \quad $\epsilon_{d} > 0$, \hspace{0.3cm} $p_{d}' = p_{d} + \epsilon_{d}$\\

\midrule

1: \textbf{Buy} all its underlying tokens from the market using $\lambda \cdot p_{d}'$ (scalar $\lambda$ depends\\
on the budget).\\

2: \textbf{Lock} all underlying tokens $u$ to \textbf{issue} $\lambda$ derivative token $d$ in the liquidity pool.\\

3: \textbf{Sell} $\lambda$ derivative token $d$ to get $\lambda \cdot p_{d}$ dollar and \textbf{reduce} derivative token price $p_{d}'$.\\

4: \textbf{Earn} profit $\lambda (p_{d}' - p_{d})$ and \textbf{repeat} steps 1 to 5 until $p_{d}' = p_{d}$, i.e. $\epsilon_{d} = 0$.\\
\bottomrule
\end{tabularx}
    \begin{tabularx}{\linewidth}{l}
\toprule

\textbf{Vault Owner's Strategy} \quad $\epsilon_{d} = 0$, \hspace{0.3cm} $\frac{\Gamma}{c_d} = \frac{\mathbf{p}_u^T\mathbf{q}_u}{c_d q_d} \le 1$, \hspace{0.3cm} $p_{d}' = p_{d} + \epsilon_{d}$\\
\midrule

1: \textbf{Observe} that the value of issued crypto-backed stablecoin, $p_d q_d$, exceeds that of \\
collateral, $p_u q_u$, in the vault. As a result, liquidators are 
disincentivized from \\
liquidating the vault.\\

1: \textbf{Sell} $q_{d}$ crypto-backed stablecoins $d$ at price $p_{d}$ in the market.\\

3: \textbf{Earn} a profit of $p_d q_d - p_u q_u$ and \textbf{decrease} the token price of $d$ by increasing its\\ supply.\\

\bottomrule
\end{tabularx}
    \label{tab:pegging_mechanism}
\end{table}
In cases where the market value of the derivative token temporarily deviates from the pegged value due to the short-term demand and supply shock ($\epsilon_{d} \neq 0$), it becomes subject to the arbitrage processes outlined in Arbitrager's Strategy 1 for market value below the pegged value ($p_{d}' < p_{d}$) and Arbitrager's Strategy 2 for market value above the pegged value ($p_{d}' > p_{d}$) in \autoref{tab:pegging_mechanism}. These two strategies will return the derivative token price to the pegged price ($p_{d}' = p_{d}$). To enhance liquidity and pegging stability, some \acp{cdp} deploy pools to lock \ac{ncb} stablecoins and issue crypto-backed stablecoins (e.g., MakerDAO's peg stability module~\cite{MakerDAO2022PegStability}). This setup enables Arbitrager's Strategies 1 and 2 to facilitate the pegging of crypto-backed stablecoins. In the case of \ac{cdp} undercollateralization where the total value of underlying tokens (i.e. collateral) is lower than that of the issued crypto-backed stablecoins $\Gamma < 1$, the derivative token depeg permanently in proportion to $\Gamma$ due to vault owners employing the Vault Owner's Strategy. Different strategies for corresponding scenarios, as outlined in \autoref{tab:pegging_mechanism}, underpin the pricing formula \autoref{eq:endo_price}.

\section{Derivation of $\Delta$ and $\Pi$}
\label{derivation_delta}

If liquidation happens in a \ac{cdp}, a liquidator will repay all debts of the position, with their value denoted as $V_d = \mathbf{d}^T \mathbf{p}_d$ according to~\autoref{subsec:liquidation}. Simultaneously, the \ac{cdp} receives and burns all repaid stablecoins (see journal entries of MakerDAO liquidation in \autoref{tab:journalization-makerdao}). Therefore, the repaid token quantity remains unchanged, hence $\Delta_{t+1} = 0$ for the repaid token. Next, the liquidator withdraws all collaterals from the \ac{cdp}, with their value denoted as $V_c= \mathbf{c}^T \mathbf{p}_c$ according to~\autoref{subsec:liquidation}. The decrease of the collateral token quantity is as follows: $\Delta_{t+1} = - q_{t}$. Before initiating the liquidation, the rational liquidator will compare the bonus $V_{c} - V_{d}$ with the gas fee $gasFees$. If the liquidation profit $\Pi = V_{c} - V_{d} - gasFees >0$, the liquidator will trigger the liquidation.

If liquidation happens in a lending protocol, a liquidator will first repay $V_{\textnormal{liq}}$ worth of debt and then receive $V_{\textnormal{liq}}(1+b)$ worth of collateral. Here, the liquidation spread $b$ represents the proportion of the bonus that a liquidator can collect relative to the debt during the liquidation. Before initiating the liquidation, the rational liquidator will compare net revenue $V_{\textnormal{liq}} b$ with the transaction fee $gasFees$. If the profit $\Pi = V_{\textnormal{liq}} b - gasFees >0$, the liquidator will trigger the liquidation. 

Theoretically, the lending protocol establishes an upper bound for $V_{\textnormal{liq}}$, defined as $\delta V_{d}$. Here, the close factor $\delta$ denotes the maximum proportion of the debt allowed to be repaid in a single liquidation. To obtain an analytic result, we assume the liquidator repays exactly this maximum amount, leading to $V_{\textnormal{liq}} = \delta V_{d}$. However, we should also consider the scenario in which the actual collateral value $V_c$ is lower than the expected collateral $V_{\textnormal{liq}}(1+b)$ that the liquidator wants to receive when preparing to liquidate the theoretical maximum amount of debt $\delta V_{d}$. Consequently, the maximum amount of debt that can be covered is determined by the lesser of $\frac{V_c}{1 + b}$ and $\delta V_{d}$, expressed as $V_{\textnormal{liq}} = \min\{\frac{V_c}{1+b}, \delta V_d\}$. Given that a position may include multiple types of collateral and borrowed tokens, we assume the liquidator withdraws and repays each proportionally to their share of the total collateral and debt, respectively. Consequently, the repaid token quantity increase by $\frac{V_{\textnormal{liq}}q}{V_d}$, while the collateral token decreases by $\frac{(1+b)V_{\textnormal{liq}}q}{V_{c}}$.

\section{Protocol-Level \ac{tvr}}
\label{ind_tvl}
\begin{table}[tb]
\centering
\caption{Account-perspective balance sheets of the Lido user and the Aave user. In the Lido scenario, the user deposits 1,000 USD \ac{eth} to receive 1,000 USD stETH and further deposits 1,000 USD stETH to generate 1,000 wstETH. The Aave scenario aligns with the process in \autoref{subfig:leverage_defi}. We highlight receivables in green and payables in red.}
\scriptsize
\begin{subtable}{0.5\linewidth}
\caption{Lido user.}
\label{subtab:acct_bs_wrapping}
\begin{tabular}{@{}p{0.55\linewidth}>{\raggedleft\arraybackslash}p{0.13\linewidth}>{\raggedleft\arraybackslash}p{0.13\linewidth}@{}}
\toprule
\textbf{Immediately After} & \encircle{2} & \encircle{4} \\
\midrule
\textit{Assets} &  \$ &  \$ \\
Receivables - stETH  & \textcolor{green}{1,000} & \textcolor{green}{0} \\
Receivables - wstETH & \textcolor{green}{-} & \textcolor{green}{1,000} \\
& \\
& \\
\midrule
\textbf{Total Assets}  & 1,000 & 1,000 \\
\bottomrule\bottomrule
\textit{Liabilities} & \$ & \$ \\
& \\
\midrule
\textbf{Total Liabilities} & 0 & 0 \\
\bottomrule
\textit{Net Positions} & \$ & \$ \\
Initial Deposit Value - ETH & 1,000 & 1,000 \\
\midrule
\textbf{Total Net Positions} & 1,000 & 1,000 \\
\bottomrule\bottomrule
\end{tabular}
\end{subtable}%
\hfill
\begin{subtable}{0.5\linewidth}
\caption{Aave user.}
\label{subtab:acct_bs_leveraging}
\begin{tabular}{@{}p{0.55\linewidth}>{\raggedleft\arraybackslash}p{0.13\linewidth}>{\raggedleft\arraybackslash}p{0.13\linewidth}>{\raggedleft\arraybackslash}p{0.13\linewidth}@{}}
\toprule
\textbf{Immediately After} & \encircle{0} & \encircle{3} & \encircle{5} \\
\midrule
\textit{Assets} & \$ & \$ & \$ \\
Cash - DAI & 2,000 & 0 & 0 \\
Cash - ETH & - & 900 & 0 \\
Receivables - aDAI & \textcolor{green}{-} & \textcolor{green}{2,000} & \textcolor{green}{2,000} \\
Receivables - aETH & \textcolor{green}{-} & \textcolor{green}{-} & \textcolor{green}{900} \\
\midrule
\textbf{Total Assets} & 2,000 & 2,900 & 2,900 \\
\bottomrule\bottomrule
\textit{Liabilities} & \$ & \$ & \$ \\
Payables - dETH & \textcolor{red}{-} & \textcolor{red}{900} & \textcolor{red}{900}\\
\midrule
\textbf{Total Liabilities}  & 0 & 0 & 0 \\
\bottomrule
\textit{Net Positions} & \$ & \$ & \$ \\
Initial Deposit Value - DAI & 2,000 & 2,000 & 2,000 \\
\midrule
\textbf{Total Net Positions}  & 2,000 & 2,000 & 2,000 \\
\bottomrule\bottomrule
\end{tabular}
\end{subtable}
\label{tab:acct_bs}
\end{table}

To avoid intra-protocol double counting, we can employ the notations in the account-perspective balance sheet to accurately evaluate the redeemable value of the individual protocol. We choose Lido as the case study to illustrate intra-protocol wrapping and Aave to show leveraging because they represent the largest protocols where these instances of double counting could occur. \autoref{subtab:acct_bs_wrapping} shows the account-perspective balance sheets of the Lido user in the intra-protocol wrapping scenario. In this scenario, a user deposits \$1k \ac{eth} (step \encircle{1}) to receive \$1k stETH in state $\mathcal{S}^{(2)}_{\textnormal{W}}$ (step \encircle{2}). Subsequently, the user deposits \$1k stETH (step \encircle{3}) to generate \$1k wstETH (step \encircle{4}). From the user's standpoint, irrespective of the frequency of intra-protocol token wrapping, the total value of receivables remains constant at \$1k in this scenario. \autoref{subtab:acct_bs_leveraging} shows the account-perspective balance sheets of the Aave user in the leveraging scenario, as illustrated in \autoref{subfig:leverage_defi}. When users take leverage to expand both protocols (see \autoref{subtab:balance-sheet-curve}) and their account's balance sheet, the total value of payables indicates the extent to which receivables are inflated, which is \$900 in the scenario. Therefore, the \ac{tvr} of an individual \ac{defi} protocol $i$ at time $t$ can be expressed as follows: ${\it TVR}_{i} = \textstyle\sum_{j \in \mathcal{A}_{i}} R_{j} -\textstyle\sum_{j \in \mathcal{A}_{i}} P_{j},$ where $P_{i}$ and $R_{i}$ denote the total payables and receivables in the balance sheet of account $i$.

\begin{figure}[tb]
\centering
\begin{subfigure}{.5\linewidth}
  \centering
  \includegraphics[width=\linewidth]{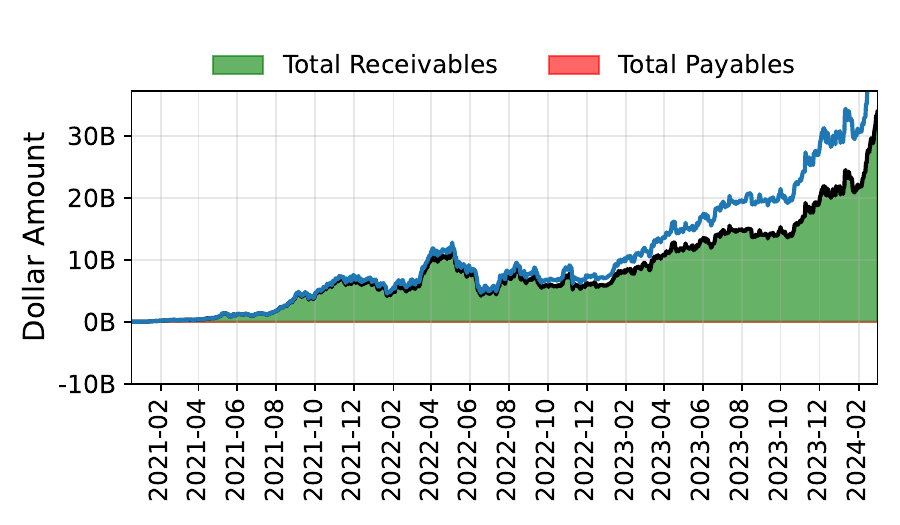}
  \caption{Lido.}
  \label{subfig:ind_lido}
\end{subfigure}%
\hfill
\begin{subfigure}{.5\linewidth}
  \centering
  \includegraphics[width=\linewidth]{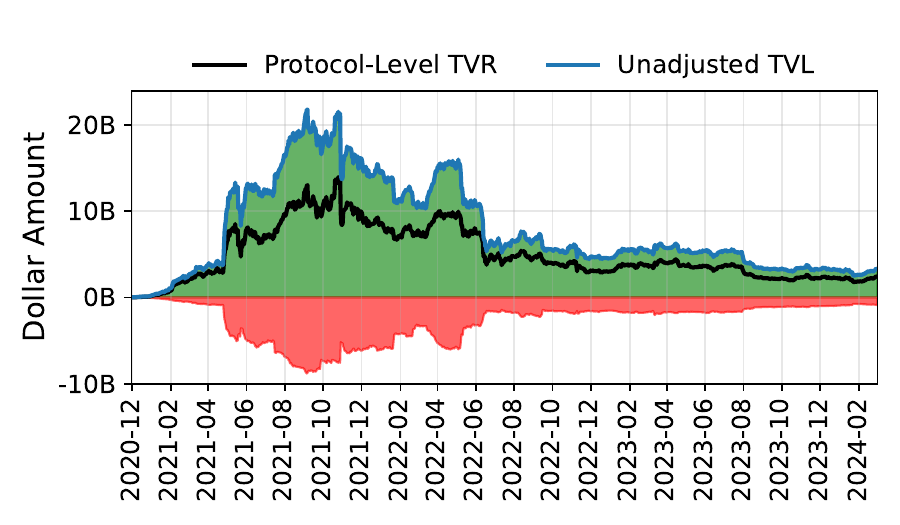}
  \caption{Aave V2.}
  \label{subfig:ind_aave}
\end{subfigure}
\caption{Total receivables, total payables, unadjusted \ac{tvl}, and protocol-level \ac{tvr} of Lido and Aave V2 in Ethereum.}
\label{fig:ind_tvl}
\end{figure}

We measure the protocol \ac{tvr} for two representative protocols Lido and Aave V2 using the on-chain data. \autoref{fig:ind_tvl} shows the total receivables, total payables, unadjusted \ac{tvl}, and protocol-level \ac{tvr} of Lido and Aave V2 in the Ethereum. The protocol \ac{tvr} of Lido is equivalent to the value of total receivables as Lido does not engage in token lending and its users do not have any payables. Users of Aave V2, a lending protocol, may have payables. Therefore, the value of these payables must be subtracted from the total receivables. We can observe the intra-protocol double counting in both Lido and Aave V2, evident from the discrepancies between the unadjusted \ac{tvl} and the protocol-level \ac{tvr}.

\section{Comparison between \ac{tvr} and DeFiLlama-adjusted 
 \ac{tvl}}
\label{compare}

\autoref{tab:compare} reports the comparison of calculation methods between DeFiLlama-adjusted \ac{defi} space ecosystem-wide \ac{tvl} and \ac{tvr}. Although MakerDAO and Convex have both plain and derivative tokens, they are classified as protocols depositing into another protocol and are therefore excluded from the DeFiLlama-adjusted \ac{tvl} calculation. In contrast, while Lido, Uniswap V2, and Curve also have derivative tokens, all their tokens are included in the DeFiLlama-adjusted \ac{tvl} calculation. As a result, the DeFiLlama-adjusted \ac{tvl} fails to accurately reflect the underlying value of \ac{defi}.

\section{Token Classification Automation}
\label{automation}









This paper acknowledges the absence of token classification automation. One possible method to automate this process and avoid reliance on third-party data is to check the opcodes of smart contracts for those tokens. For a DeFi token, one can first fetch its smart contract code to examine its \texttt{mint} functions. If the token is a derivative token, its minting requirement should depend on the availability of its underlying tokens. We provide the following three examples:
\vspace{-3mm}
\begin{enumerate}

    \item  WETH Smart Contract: To wrap ETH into WETH, the smart contract deposits ETH into its own address to mint WETH (see line~\autoref{line:weth}), as shown below:

    \begin{lstlisting}[language=Solidity, escapechar=|]
function deposit() public payable {
    balanceOf[msg.sender] += msg.value;
    Deposit(msg.sender, msg.value); |\label{line:weth}|
}\end{lstlisting}

    \newpage

    \item stETH in Lido: To wrap ETH into stETH, the Lido smart contract first checks whether deposits (ETH) are available before proceeding (see line~\autoref{line:steth}). The relevant code snippet is shown below:

    \begin{lstlisting}[language=Solidity, escapechar=|]
function deposit(uint256 _maxDepositsCount, uint256 _stakingModuleId, bytes _depositCalldata) external {
    ILidoLocator locator = getLidoLocator();

    require(msg.sender == locator.depositSecurityModule(), "APP_AUTH_DSM_FAILED");
    require(canDeposit(), "CAN_NOT_DEPOSIT");

    IStakingRouter stakingRouter = _stakingRouter();
    uint256 depositsCount = Math256.min(
        _maxDepositsCount,
        stakingRouter.getStakingModuleMaxDepositsCount(_stakingModuleId, getDepositableEther())
    );

    uint256 depositsValue;
    if (depositsCount > 0) {|\label{line:steth}|
        depositsValue = depositsCount.mul(DEPOSIT_SIZE);
        /// @dev firstly update the local state of the contract to prevent a reentrancy attack,
        ///     even if the StakingRouter is a trusted contract.
        BUFFERED_ETHER_POSITION.setStorageUint256(_getBufferedEther().sub(depositsValue));
        emit Unbuffered(depositsValue);

        uint256 newDepositedValidators = DEPOSITED_VALIDATORS_POSITION.getStorageUint256().add(depositsCount);
        DEPOSITED_VALIDATORS_POSITION.setStorageUint256(newDepositedValidators);
        emit DepositedValidatorsChanged(newDepositedValidators);
    }

    /// @dev transfer ether to StakingRouter and make a deposit at the same time. All the ether
    ///     sent to StakingRouter is counted as deposited. If StakingRouter can't deposit all
    ///     passed ether it MUST revert the whole transaction (never happens in normal circumstances)
    stakingRouter.deposit.value(depositsValue)(depositsCount, _stakingModuleId, _depositCalldata);
}\end{lstlisting}

    \item  aTokens in Aave V2: To wrap plain tokens into aTokens on Aave V2, the reserve token is transferred from the user's wallet to the liquidity pool (see line~\autoref{line:atoken}). The relevant smart contract logic is as follows:

    \begin{lstlisting}[language=Solidity, escapechar=|]
function deposit(address _reserve, uint256 _amount, uint16 _referralCode)
    external
    payable
    nonReentrant
    onlyActiveReserve(_reserve)
    onlyUnfreezedReserve(_reserve)
    onlyAmountGreaterThanZero(_amount)
{
    AToken aToken = AToken(core.getReserveATokenAddress(_reserve));

    bool isFirstDeposit = aToken.balanceOf(msg.sender) == 0;

    core.updateStateOnDeposit(_reserve, msg.sender, _amount, isFirstDeposit);

    //minting AToken to user 1:1 with the specific exchange rate
    aToken.mintOnDeposit(msg.sender, _amount);

    //transfer to the core contract
    core.transferToReserve.value(msg.value)(_reserve, msg.sender, _amount);|\label{line:atoken}|

    //solium-disable-next-line
    emit Deposit(_reserve, msg.sender, _amount, _referralCode, block.timestamp);
}\end{lstlisting}

\end{enumerate}

Therefore, we then check whether the \texttt{mint} function of this token is based on the existence of other tokens. If it is not, it should be a plain token. In the Ethereum opcode level, the smart contracts of derivative tokens without ETH as underlying always involve the operation of calling the \texttt{transferFrom} function of other ERC-20 tokens. These operations are associated with the opcodes \texttt{CALL}, \texttt{PUSH4 0x23b872dd}. \texttt{CALL} is the opcode to call a method in another contract, while \texttt{PUSH4 0x23b872dd} is the function selector \texttt{transferFrom}. On the other hand, the smart contracts of derivative tokens with ETH as underlying always involve the operation of transferring the ETH. This operation is associated with the opcode \texttt{CALLVALUE}. By identifying those opcodes, we can effectively distinguish the derivative tokens and plain tokens. Future research could explore automation techniques for distinguishing plain and derivative tokens.

The automation methods mentioned above still face several challenges and limitations. Different blockchains have distinct sets of opcodes, requiring tailored automation methods for each blockchain. Additionally, some smart contracts are upgradable, meaning their rules can change over time. Consequently, automation methods must also be updated to remain effective. Also note that not all token smart contracts follow standard implementations and some have custom designs, making it difficult for automation methods to cover every possible contract.

\begin{table}[tb]
    \centering
    \tiny
    \caption{The comparison of calculation processes between the \ac{defi} space ecosystem-wide DeFiLlama-adjusted \ac{tvl} and \ac{tvr}. The \ac{tvr} contains only a single column because it is calculated by aggregating the value of all eligible tokens. In contrast, the DeFiLlama-adjusted \ac{tvl} includes protocol-specific columns, as it is calculated by first aggregating the \ac{tvl} of eligible protocols within a blockchain and then summing the \ac{tvl} across all blockchains (see \autoref{subsec:tvl_tvr}).}
    \label{tab:compare}
    \begin{tabularx}{\linewidth}{>{\raggedright\arraybackslash}X>{\centering\arraybackslash}X>{\centering\arraybackslash}X>{\centering\arraybackslash}X>{\centering\arraybackslash}X>{\centering\arraybackslash}X>{\centering\arraybackslash}X>{\centering\arraybackslash}X}
    \toprule
    \multirow{3}{*}{\textbf{\makecell[c]{Tokens}}} & \multirow{3}{*}{\textbf{TVR}} & \multicolumn{6}{c}{\textbf{DeFiLlama-adjusted TVL}}\\
    \cline{3-8}
    && \multirow{2}{*}{\textbf{Lido}} & \multirow{2}{*}{\textbf{MakerDAO}} & \multirow{2}{*}{\textbf{\makecell{Aave\\V2}}} & \multirow{2}{*}{\textbf{\makecell{Uniswap\\V2}}} & \multirow{2}{*}{\textbf{Curve}} & \multirow{2}{*}{\textbf{Convex}} \\
    &&&&&&&\\
    \midrule
    ETH  & \CIRCLE & \CIRCLE & \Circle & \CIRCLE & \CIRCLE & \CIRCLE & -\\
    WETH & \Circle & \CIRCLE & \Circle & \CIRCLE & \CIRCLE & \CIRCLE & -\\
    wstETH & \CIRCLE & - & \Circle & - & \CIRCLE & \CIRCLE & -\\
    USDT & \CIRCLE & - & \Circle & \CIRCLE & \CIRCLE & \CIRCLE & -\\
    USDC & \CIRCLE & - & \Circle & \CIRCLE & \CIRCLE & \CIRCLE & -\\
    WBTC & \Circle & - & \Circle & \CIRCLE & \CIRCLE & \CIRCLE & -\\
    DAI & \Circle & - & \Circle & \CIRCLE & \CIRCLE & \CIRCLE & -\\
    a3CRV & \Circle & - & - & - & \CIRCLE & - & \Circle\\
    USDP & \CIRCLE & - & \Circle & \CIRCLE & \CIRCLE & \CIRCLE & -\\
    AAVE & \CIRCLE & - & - & \CIRCLE & \CIRCLE & \CIRCLE & -\\
    \bottomrule
    \multicolumn{8}{l}{$\CIRCLE$: The token value is included in the metric calculation.}\\
    \multicolumn{8}{l}{$\Circle$: The token value is not included in the metric calculation.}\\
    \multicolumn{8}{l}{$-$: The absence of an asset-specific liquidity pool in the respective protocol.}\\
\end{tabularx}
\end{table}

\section{Environment Settings for Sensitivity Tests}
\label{sec:params}
\autoref{tab:params} shows the environment settings for sensitivity tests.

\begin{table}[b]
\centering
\tiny
\caption{Environment settings for sensitivity tests, including parameter details and \ac{plf} position information parsed from on-chain data.}
\begin{tabular}
{>{\centering\arraybackslash}m{2cm}>{\centering\arraybackslash}m{1.5cm}>{\centering\arraybackslash}m{2.8cm}>{\centering\arraybackslash}m{2.8cm}>{\centering\arraybackslash}m{2.8cm}}

\toprule

\multirow{3}{*}{Parameters}  & \multirow{3}{*}{\makecell{Related\\Equation}}  & \multicolumn{3}{c}{Value}\\ 

\cline{3-5} 
 &&&&\\
 & & Maximum TVL (2021-12-02) & LUNA collapse (2022-05-09) & FTX collapse (2022-11-08)\\
 \midrule
\href{https://etherscan.io/chart/etherprice}{$p_{\textnormal{ETH}}$} & \autoref{eq:endo_price}, \autoref{eq:endo_quantity} &\$4,075.03 & \$2,249.89 & \$1,334.29\\
\href{https://aave.com/docs#calculating-profitability-vs-gas-cost}{$\delta_{\textnormal{AAVE}}$} & \autoref{eq:endo_quantity} & 0.5 & 0.5 & 0.5\\
\href{https://docs.makerdao.com/smart-contract-modules/dog-and-clipper-detailed-documentation}{$\delta_{\textnormal{MKR}}$} & \autoref{eq:endo_quantity} & 1 & 1 & 1\\
\href{https://etherscan.io/chart/gasprice}{$gasPrice$} & \autoref{eq:endo_quantity} & $\$4.85 \times 10^{-4}$ & $\$1.50 \times 10^{-4}$ & $\$7.60 \times 10^{-5}$\\
$gasLimit$~\cite{Tovanich2023ContagionCompound} & \autoref{eq:endo_quantity} & $500,000$ & $500,000$ & $500,000$\\
\bottomrule
\multirow{3}{*}{PLF} & \multirow{3}{*}{\makecell{Related\\Equation}} & \multicolumn{3}{c}{Number of accounts}\\ 

\cline{3-5} 
 &&&&\\
 & & Maximum TVL (2021-12-02) & LUNA collapse (2022-05-09) & FTX collapse (2022-11-08)\\
 \midrule
MakerDAO & \autoref{eq:endo_quantity} & 27,051 & 28,220 & 29,711\\
Aave & \autoref{eq:endo_quantity} & 21,336 & 30,859 & 47,957\\
\bottomrule
\end{tabular}
\label{tab:params}
\end{table}

\begin{figure}[tb]
\centering
\begin{subfigure}{0.47\linewidth}
    \includegraphics[width=\linewidth]{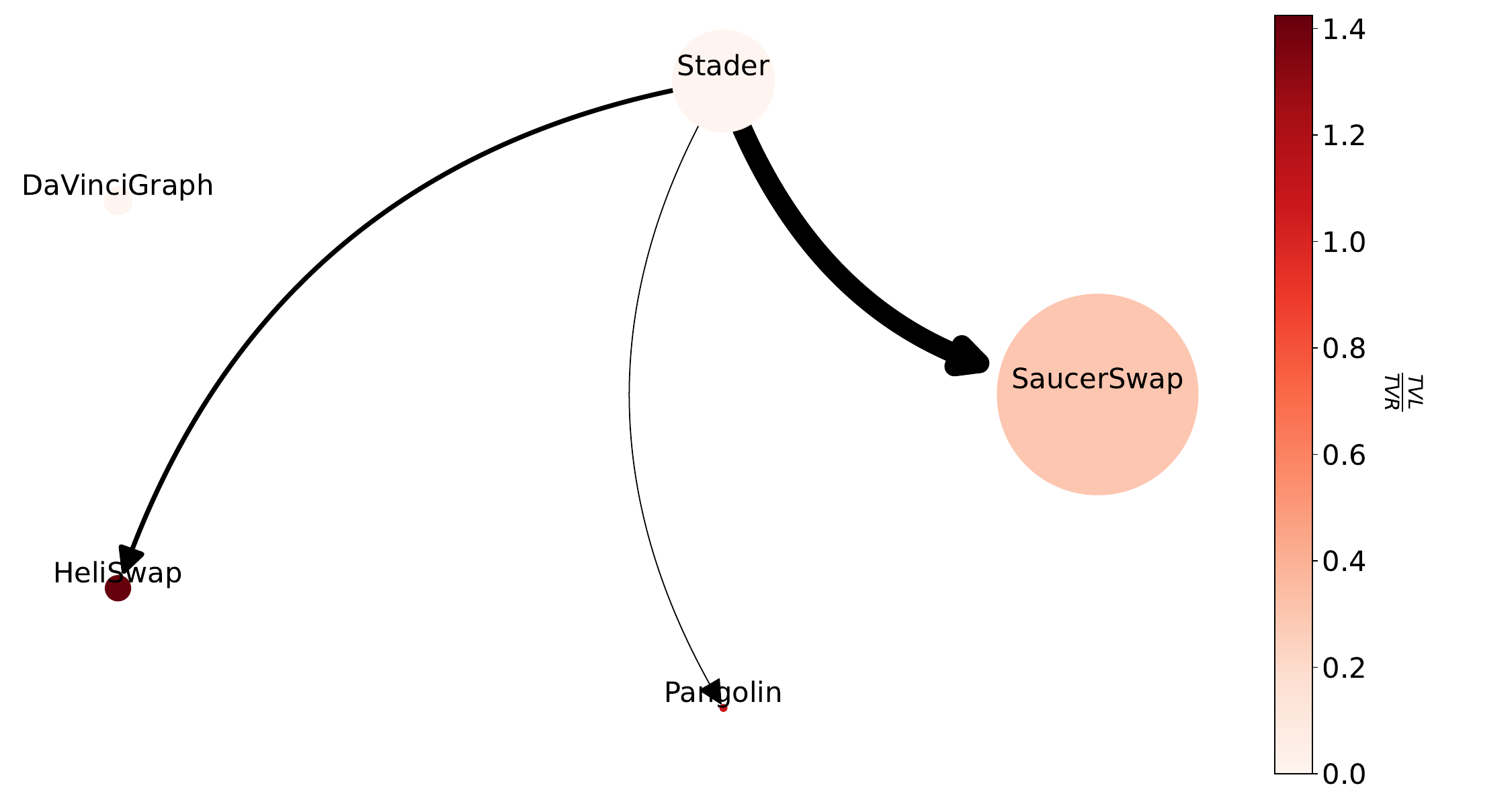}
    \caption{Token wrapping network of Hedera on 2024-03-01. Node size corresponds to the \ac{tvl}, edge width represents the dollar amount of tokens, and node color reflects the ratio between \ac{tvl} and \ac{tvr}.}
    \label{fig:hedera_network}
\end{subfigure}%
\hfill
\begin{subfigure}{0.47\linewidth}
    \includegraphics[width=\linewidth]{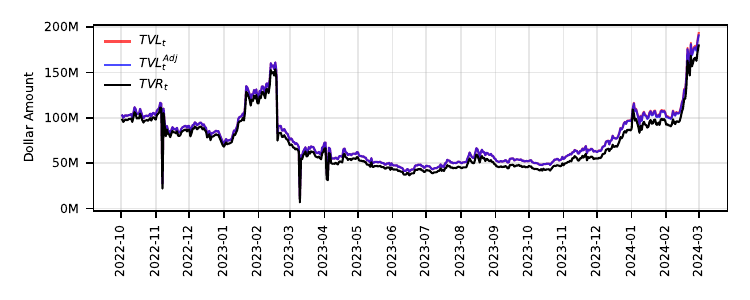}
    \caption{\ac{tvl} and \ac{tvr} of Hedera over time, where the red, blue, and black lines represent the DeFiLlama \ac{tvl} subjected to double counting, DeFillama-adjusted \ac{tvl}, and \ac{tvr}.}
    \label{fig:tvl_hedera}
\end{subfigure}
    \caption{Hedera double counting analysis.}
\end{figure}

\section{Altchain Analyses}
\label{altchain_analyses}

To gain a broader understanding of the double-counting issue, we analyze two niche \acp{altchain}: Hedera ranked 42nd in \ac{tvl} and Ultron ranked 21st in \ac{tvl}. Our findings indicate that blockchains with limited infrastructure feature simpler token-wrapping networks, which leads to less double counting. Additionally, we observe that the DeFiLlama framework addressing the double counting might deflate the true value redeemable of an altchain.

\subsection{Hedera}

Hedera is a public hashgraph blockchain and governing body tailored to meet the requirements of mainstream markets~\cite{Baird2018Hedera:Council}. As shown in \autoref{fig:hedera_network}, the token wrapping network of Hedera is relatively simple. Hedera has only eight \ac{defi} protocols, three of which have zero \ac{tvl} and have been shut down. Among the other six protocols, only one liquidity staking protocol named Stader can generate receipt tokens that can be subsequently deposited into other protocols. Compared to the networks in Ethereum mentioned in \autoref{sec:risk_analysis}, Hedera has a simpler token-wrapping network, thus experiencing less double counting under the \ac{tvl} framework, as shown in \autoref{fig:tvl_hedera}. We can also partially attribute hedera's low \ac{tvl}-\ac{tvr} ratio to the fact that its native coin HBAR also has token-like functions, rendering wrapping of HBAR unnecessary~\cite{Hedera2024SendContracts}. 

\begin{figure}[tb]
    \centering
    \begin{subfigure}{0.47\linewidth}
        \includegraphics[width=\linewidth]{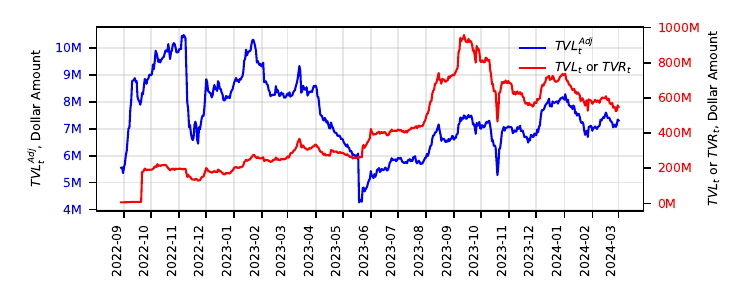}
        \caption{TVL and TVR of Ultron over time. The left-hand-side y-axis denotes the dollar amount of DeFillama-adjusted TVL, while the right-hand-side y-axis represents the DeFiLlama TVL subjected to double counting.}
        \label{fig:tvl_ultron}
    \end{subfigure}%
    \hfill
    \begin{subfigure}{0.47\linewidth}
        \includegraphics[width=\linewidth]{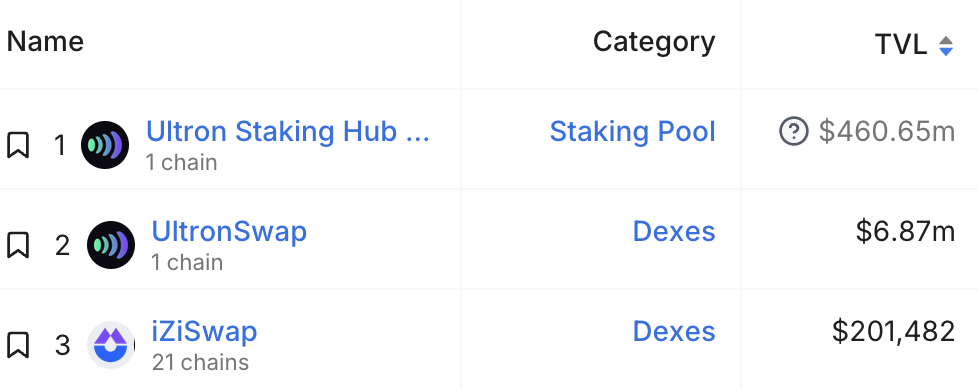}
        \caption{DeFiLlama removes the TVL of the Ultron Staking Hub NFT protocol from Ultron's TVL excluding double counting, as this protocol falls under the category of protocols that deposit into another protocol.}
      \label{fig:double-count-removal-ultron}
    \end{subfigure}
    \caption{Ultron double counting analysis.}
\end{figure}

\subsection{Ultron}

Ultron is a layer-one blockchain~\cite{Ultron2023UltronWhitepaper} with three \ac{defi} protocols in its ecosystem. The Ultron Staking Hub NFT, created by the Ultron Foundation, serves as a digital asset growth instrument allowing users to earn daily \ac{apr} returns in Ultron native tokens. Each user can mint \acp{nft} and stake them on the protocol for 5 years with a vesting schedule. All liquidity is securely locked within a staking smart contract and can be claimed at specific timelines. UltronSwap and iZiSwap are two \acp{dex} on Ultron.
Ultron does not involve wrapping or leverage. \Acp{nft} minted by Ultron Staking Hub NFT and \ac{lp} tokens generated by two \acp{dex} cannot be further deposited into other protocols. Therefore, Ultron is not subject to the double counting problem under the \ac{tvr} framework. However, DeFiLlama removes the Ultron Staking Hub NFT protocol from its adjusted \ac{tvl}, falsefully suggesting this protocol deposits into another protocol. The inaccuracy of the methodology (see \autoref{subsec:tvl}) makes the DeFillama-adjusted TVL of Ultron significantly lower than its \ac{tvr} as shown in \autoref{fig:tvl_ultron}.

\begin{table}[b]
    \centering
    \tiny
    \renewcommand{\arraystretch}{0.5}
    \caption{Spearman's rank correlation coefficients \cite{Spearman1904TheThings} between the natural logarithmic returns of macroeconomic indicators, cryptocurrency market indicators, and \ac{defi} money multiplier computed from \ac{tvl} and \ac{tvr}.}
    \begin{tabularx}{\linewidth}{>{\raggedright\arraybackslash}p{1.5cm}*{6}{X}}
\toprule
  & \multicolumn{3}{c}{Macroeconomic / TradFi indicators} & \multicolumn{3}{c}{Cryptocurrency / DeFi indicators}\\
 \cmidrule(lr){2-4}\cmidrule(lr){5-7}
 & $\ln \frac{CPI_t}{CPI_{t-1}}$ & $\ln \frac{VIX_t}{VIX_{t-1}}$ & $\ln \frac{M_t^{\textnormal{TradFi}}}{M_{t-1}^{\textnormal{TradFi}}}$ & $\ln \frac{ETH_t}{ETH_{t-1}}$ & $\ln \frac{S\&P_t}{S\&P_{t-1}}$ & $\ln \frac{M_t^{\textnormal{DeFi}}}{M_{t-1}^{\textnormal{DeFi}}}$ \\
$\ln \frac{CPI_t}{CPI_{t-1}}$ & {\cellcolor[HTML]{FF8080}} \color[HTML]{F1F1F1} 1.00$^{***}$ & {\cellcolor[HTML]{F0F0FF}} \color[HTML]{000000} -0.11 & {\cellcolor[HTML]{FFCCCC}} \color[HTML]{000000} 0.40$^{**}$ & {\cellcolor[HTML]{FFF8F8}} \color[HTML]{000000} 0.06 & {\cellcolor[HTML]{FFFAFA}} \color[HTML]{000000} 0.04 & {\cellcolor[HTML]{FEFEFF}} \color[HTML]{000000} -0.01 \\
$\ln \frac{VIX_t}{VIX_{t-1}}$ & {\cellcolor[HTML]{F0F0FF}} \color[HTML]{000000} -0.11  & {\cellcolor[HTML]{FF8080}} \color[HTML]{F1F1F1} 1.00$^{***}$ & {\cellcolor[HTML]{E0E0FF}} \color[HTML]{000000} -0.25 & {\cellcolor[HTML]{E8E8FF}} \color[HTML]{000000} -0.17 & {\cellcolor[HTML]{E0E0FF}} \color[HTML]{000000} -0.25 & {\cellcolor[HTML]{FFFCFC}} \color[HTML]{000000} 0.02 \\
$\ln \frac{M_t^{\textnormal{TradFi}}}{M_{t-1}^{\textnormal{TradFi}}}$ & {\cellcolor[HTML]{FFCCCC}} \color[HTML]{000000} 0.40$^{**}$ & {\cellcolor[HTML]{E0E0FF}} \color[HTML]{000000} -0.25 & {\cellcolor[HTML]{FF8080}} \color[HTML]{F1F1F1} 1.00$^{***}$ & {\cellcolor[HTML]{FAFAFF}} \color[HTML]{000000} -0.04 & {\cellcolor[HTML]{F6F6FF}} \color[HTML]{000000} -0.06 & {\cellcolor[HTML]{F6F6FF}} \color[HTML]{000000} -0.08 \\
\bottomrule
\end{tabularx}
    \raggedright ***, **, and * denote the 1\%, 5\%, and 10\% significance levels, respectively.
    \label{fig:log_corr}
\end{table}
\section{Robustness tests for Correlation Coefficients}
\label{sec:log_corr}

As a robustness test for~\autoref{tab:corr}, we also calculate Spearman’s rank correlation coefficients~\cite{Spearman1904TheThings} between the natural logarithmic return of these indicators to make variables stationary, which is shown in \autoref{fig:log_corr}.





\end{document}